*Article*

# Forecasting the Intra-Day Spread Densities of Electricity Prices

**Ekaterina Abramova \*,† and Derek Bunn †**

London Business School, Department of Management Science and Operations, Regent's Park,
London NW1 4SA, UK; dbunn@london.edu
\* Correspondence: eabramova@london.edu
† These authors contributed equally to this work.

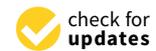

**Abstract:** Intra-day price spreads are of interest to electricity traders, storage and electric vehicle operators. This paper formulates dynamic density functions, based upon skewed-t and similar representations, to model and forecast the German electricity price spreads between different hours of the day, as revealed in the day-ahead auctions. The four specifications of the density functions are dynamic and conditional upon exogenous drivers, thereby permitting the location, scale and shape parameters of the densities to respond hourly to such factors as weather and demand forecasts. The best fitting and forecasting specifications for each spread are selected based on the Pinball Loss function, following the closed-form analytical solutions of the cumulative distribution functions.

## 1. Introduction

Whilst day ahead electricity price forecasting has been a topic of substantial and wide-ranging research in terms of methods, the focus has mostly been upon price levels for the delivery periods (usually hourly) the following day. More recently there has been an interest in density forecasts for the hourly prices, motivated by considerations of risk management. See [1,2] for extensive reviews. In this paper, we provide a new formulation with a focus upon price spreads, and specifically, we forecast the density functions for the intraday spreads in the day-ahead prices. The optimal operation of storage facilities, e.g., batteries and electric vehicles, or load shifting programmes, e.g., demand-side management, over daily cycles depends upon these spreads if they are operated as merchants, arbitraging buying and selling from the wholesale market. Furthermore, if the risk is a consideration, analysis of the mean differences in price levels would be inadequate, and we therefore directly estimate the density functions of all hourly spreads in prices at the day-ahead stage. These forecasts ahead of the day-ahead auctions would be needed to help traders decide whether they want to be buyers or sellers at each hour and thereby optimise their bids and offers. Our specification, estimation and forecasting of these arbitrage spreads are new and computationally-intensive.

Based upon day-ahead forecasts for the drivers of electricity prices, such as demand, wind and solar production, gas and coal prices, forecasts for electricity price levels have been proposed from various methods, e.g., [3–6] and some for price densities [1,7], but apparently no methods have been developed specifically for forecasting intraday spread densities. Until recently storage assets, such as pumped hydro storage, would regularly store energy at night and discharge at the daily peak demand periods, which were quite predictable. However with the penetration of wind and especially solar generating facilities, the peak and trough hours in prices move around the day and in sunny locations with substantial solar energy, e.g., California, the lowest prices may often be in the middle of the day [8]. Thus, the expected daily spreads in prices, and their consequent arbitrage opportunities for storage or load shifting, will depend upon the wind and solar forecasts, as well as demand and supply considerations. Furthermore, the price density functions are non-normal with skewness switching between positive and negative depending upon the dynamics of production of renewable energy [9]. Because of this non-Normality and the non-independence of each hourly price, spread densities cannot be easily derived as the difference of the price densities, and so we estimate and forecast the daily matrix of intraday spreads directly.

We apply our spread densities formulation to the German market. This is the largest and the main daily reference market for wholesale power in Europe. It is also strongly influenced by wind and solar production, as well as providing a context where batteries and demand-side management are active innovations. The day ahead auction has been actively researched and closes at noon each day, with the vector of 24 hourly prices for the next day being released an hour later. For modeling the spread densities, we adapt the Generalised Additive Model for Location, Scale and Shape (GAMLSS) semi-parametric regression model [10], which has already been used effectively to form day-ahead densities of price levels in the German context [9]. Within this framework, the hourly electricity price spreads form a response variable, whose distribution function varies according to multiple exogenous factors. The GAMLSS framework allows choice from a wide range of distributions, whose distribution parameters change according to the exogenous variables specified using (non)linear relationships. The dynamic location, scale and shape parameters (related to the mean, volatility, skewness and kurtosis of price spreads) are therefore explicitly incorporated into the forecasting model.

The paper proceeds by first describing the data and the density estimation process. In Section 2, we use the Pinball Loss function to select the best fitting density model with four distribution parameters. Then in Section 3, we undertake a rolling window forecasting evaluation and demonstrate the value of the dynamic, conditional latent parameter. Section 4 concludes.

## 2. Data and Methodology

### 2.1. Data

The German hourly day-ahead electricity price, wind forecast, solar forecast and actual total load data were downloaded from the Open Power System Data website https://data.open-power-system-data.org for the period of 1 January 2012 to 31 March 2017 (resulting in $t = 1, ..., 1917$ time steps). The data is comprised of four German control areas in MW: 50 Hertz, Amprion, TenneT and TransnetBW. The summer time hour change was accounted for by creating hour 02 with interpolation between hours 01 and 03. For the clock change back the (later) 02 hour was deleted. The German day-ahead total load data is calculated as an average of $4 \times 15$ min segments following the beginning of each hour. The "actual total load" for each control area (obtained at the end of the 15 min real bidding time in the balancing markets) is averaged and results across four regions are summed up to give the total actual load. We use the steam coal ARA 1 month forward benchmark index for steam coal (one price per month) and the Germany Gaspool (GPL) natural gas day-ahead forward (one price per day) for gas. The weekly seasonality and holidays are included into a single dummy variable which takes on value 1 for Saturday/Sunday and the following German state holidays: New Year's Day, Good Friday, Easter Monday, Labour Day, Ascension Day, Whit Monday, German Unity Day, Christmas Day, Boxing Day and New Years Eve (see Appendix A for further details).

Day ahead electricity price data are plotted for a selection of hours, showing the evolution of fitted distribution over the year 2017. The distribution of choice 'ST5' (to be described later) was used to fit the histogram data from the GAMLSS functions library. The hourly electricity spot price data display high volatility, fast-changing dynamics and highly skewed distributions (see Figure 1).

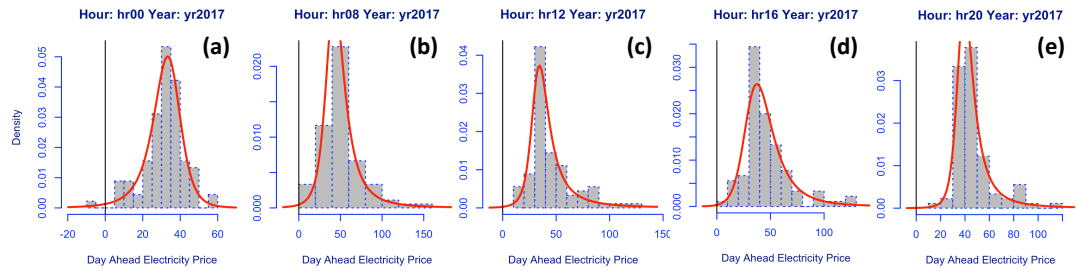

**Figure 1.** Fitted German electricity prices for year 2017 for hours: (**a**) midnight, (**b**) 8 a.m., (**c**) noon, (**d**) 4 p.m., (**e**) 8 p.m. (vertical black line shows price €0).

The day ahead electricity price data also shows moderate/strong correlations between prices at different hour of the day, as depicted by highly-positive off-diagonal entries in Figure 2 (note the smallest value is 0.41).

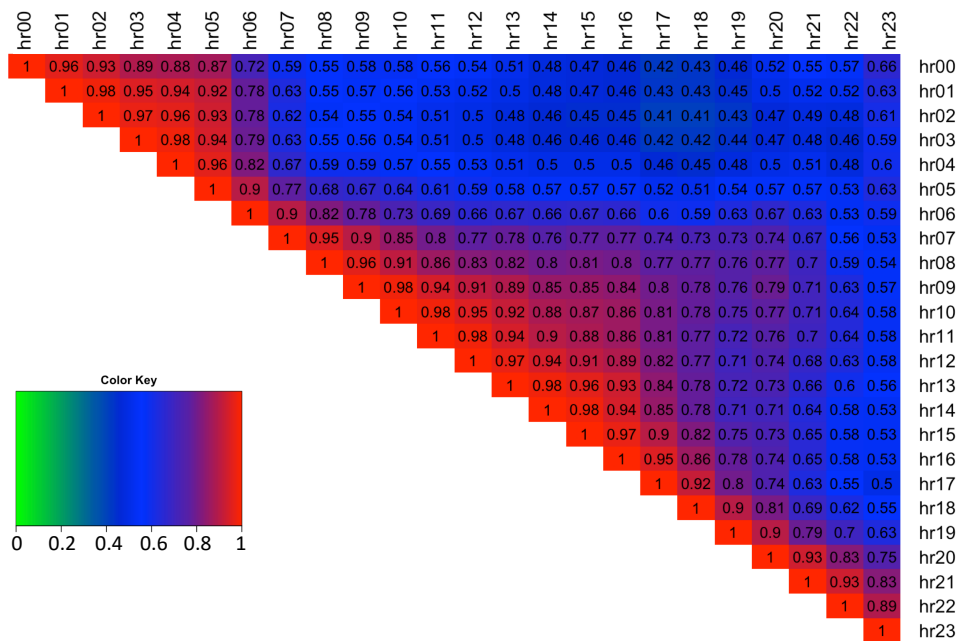

**Figure 2.** Correlation matrix of intraday electricity prices.

### 2.2. Model Description

An intraday spread, $Y_t^{(s)}$, between two hours, denoted by spread number $s$, of the day-ahead electricity spot prices, is calculated by taking away the later hour price information from the earlier one, resulting in a positive spread if later hour is less expensive and negative otherwise. Each day therefore comprises of 276 separate spreads, and with 1917 days in the time series, the full data set is denoted by $\mathbf{Y} \in \mathbb{R}^{1917 \times 276}$. Each electricity spread time series is tested for stationarity using the Augmented Dickey–Fuller test (`adf.test()` with 10 lags and a linear trend) and is confirmed to be stationary at the 1% significance level.

We plot example histograms for intraday spreads obtained between hours 00–08, 08–12, 12–16, 16–20 for the example year 2017. Figure 3 depicts the high skewness and kurtosis of the spreads and shows the variation between spread distribution shapes for different hours of the day. Note data is symmetrical about the diagonal thus only the upper diagonal is plotted.

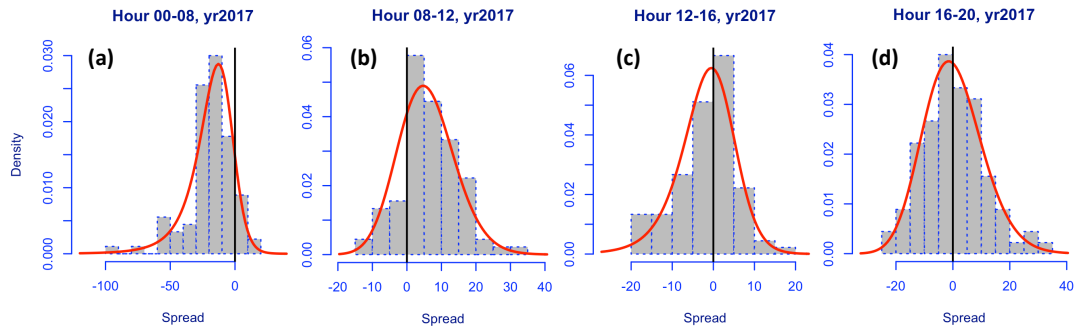

**Figure 3.** Fitted German electricity spread prices for year 2017 for hours: (**a**) 00–08 a.m.; (**b**) 08–12 noon; (**c**) 12–16pm; (**d**) 16–20 p.m. (vertical black line shows price €0).

The hourly spread data possesses a high degree of skewness in the range $[-8.31, 7.85]$ (see Figure 4). The plot shows that typically the spreads are negatively skewed (note spreads are obtained by $hr_{earlier} - hr_{later}$).

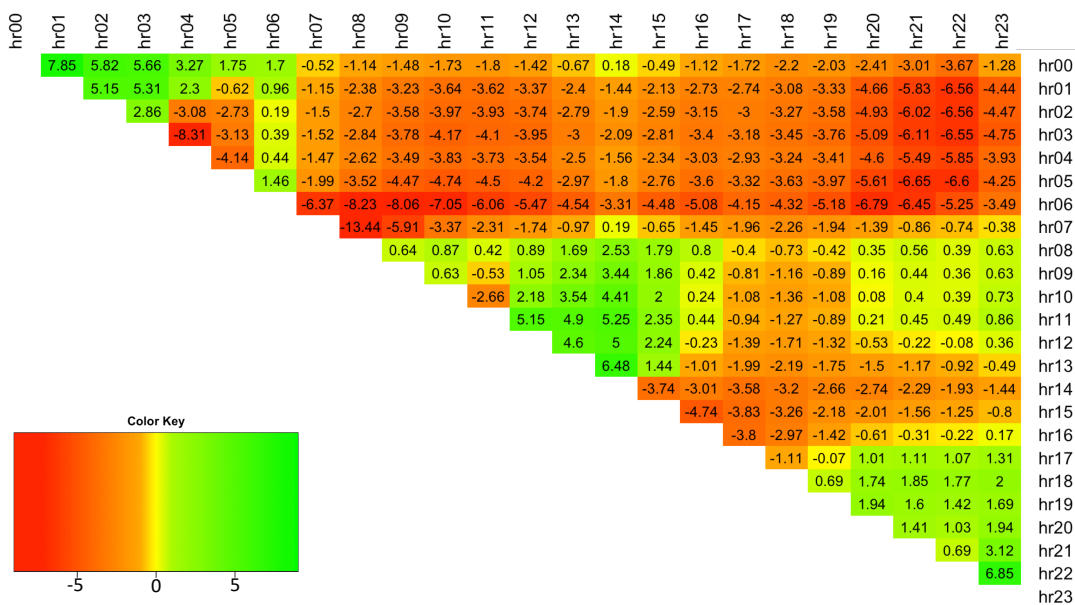

**Figure 4.** Skewness for each electricity spread.

The hourly spread data displays a high degree of positive excess kurtosis and thus indicates that spreads have Leptokurtic distributions in the range of $[3.8, 140]$ (see Figure 5).

Therefore, given the above analysis, we chose to model the full four-parameter distribution of electricity price spreads at each time step (day), $t$. The spreads between exogenous variables are likewise formed by taking away the later hour values from the earlier ones in all cases except for the gas forward prices, coal forward prices and the dummy variable, for which only the daily, rather than hourly, values are available. Hence the following exogenous variables are considered for modeling $Y_t^{(s)}$: (1) spread of the lagged intraday electricity price, (2) gas Gaspool forward daily price, (3) coal ARA forward daily price, (4) spread of wind day-ahead forecast, (5) spread of solar day-ahead forecast, (6) dummy variable taking the value of 1 for weekends/holidays, (7) spread of the day-ahead total load forecast, and (8) an interaction load variable, calculated as $Load_{spread} * \frac{1}{2} (Load_{earlierHr} + Load_{laterHr}) = \frac{1}{2} * (Load^2_{earlierHr} - Load^2_{laterHr})$ i.e., an average of the load for the two hours from which the spread is calculated, weighted by the load spread obtained for those hours. This variable provides interaction of $\Delta \ Load * avLoad$ in order to account for the rate of change in load. The full exogenous variables spread data is a 3D matrix denoted by $\mathbf{X} \in \mathbb{R}^{1917 \times 9 \times 276}$, where the first column is a vector of 1s used in the calculation of the off-set.

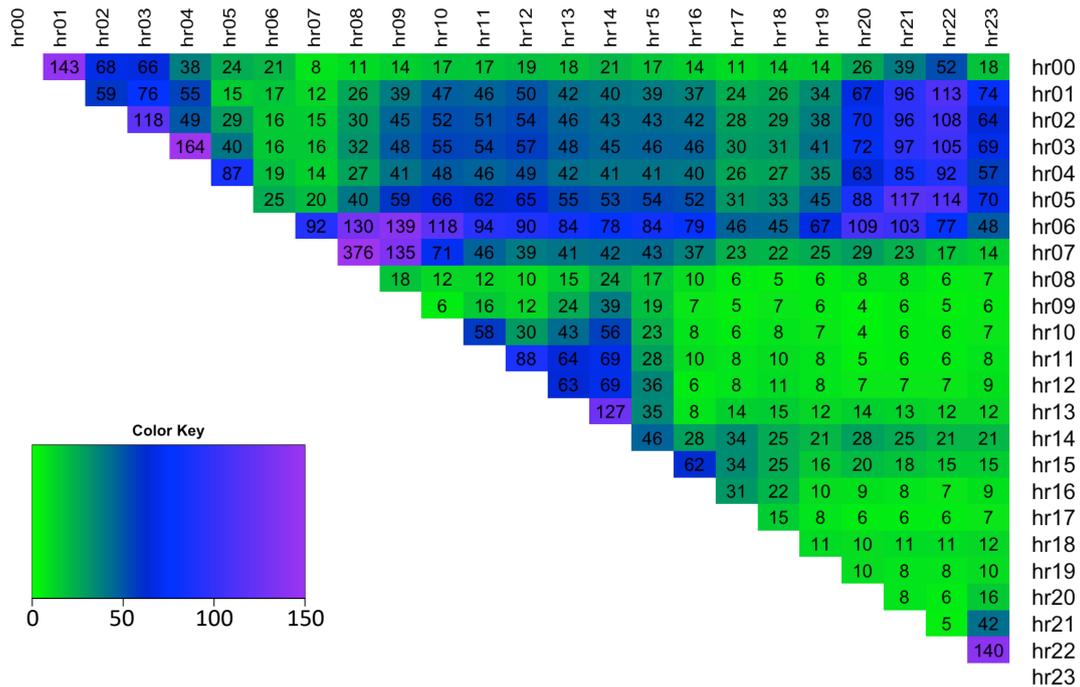

**Figure 5.** Excess kurtosis for each electricity spread.

Estimation using the Generalised Additive Model for Location, Scale and Shape (GAMLSS) framework [11] facilitates modeling each parameter ($\mu, \sigma, \nu, \tau$) of the response variable's distribution as a function of explanatory variables. The range of possible distributions includes both exponential family (as per Generalised Linear Model (GLM)) and general family distributions, which allow for both discrete/continuous and highly skewed/kurtotic distributions (unlike GLM). The probability density function of response variable for $T$ observations is given by $f_Y(y_t^{(s)}|\boldsymbol{\theta}_t^{(s)}) \sim D(\boldsymbol{\theta}_t^{(s)})$, where $s = 1, \ldots, 276$ indicates a number given for each unique spread between two intra-day hours, $\boldsymbol{\theta}_t^{(s)} = [\theta_{t,1}^{(s)}, \theta_{t,2}^{(s)}, \theta_{t,3}^{(s)}, \theta_{t,4}^{(s)}]^T = [\mu_t^{(s)}, \sigma_t^{(s)}, \nu_t^{(s)}, \tau_t^{(s)}]^T$ is a vector of distribution parameters, and $D(\boldsymbol{\theta}_t^{(s)})$ represents the distribution of spread number $s$ on day $t$. We use parametric linear GAMLSS framework which relates distribution parameters to explanatory variables by

$$g_k(\boldsymbol{\theta}_k^{(s)}) = \boldsymbol{\eta}_k^{(s)} = \mathbf{X}_k^{(s)}\boldsymbol{\beta}_k^{(s)} \tag{1}$$

where $k = 1, 2, 3, 4$ specifies the distribution parameter corresponding to $\mu, \sigma, \nu, \tau$ respectively; $\boldsymbol{\theta}_k^{(s)} \in \mathbb{R}^T$ is a vector comprised of values for distribution parameter $k$ over $t = 1, \ldots, T$ time steps (i.e., $\boldsymbol{\theta}_1^{(s)} = \boldsymbol{\mu}^{(s)}, \boldsymbol{\theta}_2^{(s)} = \boldsymbol{\sigma}^{(s)}, \boldsymbol{\theta}_3^{(s)} = \boldsymbol{\nu}^{(s)}, \boldsymbol{\theta}_4^{(s)} = \boldsymbol{\tau}^{(s)}$); $T$ is the number of observations; $g_k(\cdot)$ is the monotonic link function of distribution parameter $k$; $\boldsymbol{\eta}_k^{(s)} \in \mathbb{R}^T$ is the linear predictor vector for distribution parameter $k$; $\boldsymbol{\beta}_k^{(s)} = [\beta_{0,k}^{(s)}, \beta_{1,k}^{(s)}, \ldots, \beta_{J_k,k}^{(s)}]^T \in \mathbb{R}^{J_k+1}$ vector of coefficients learnt for parameter $k$; $J_k$ is the number of significant exogenous variables for parameter $k$ obtained at 5% significance level; and $\mathbf{X}_k^{(s)} \in \mathbb{R}^{T \times J_k+1}$ is the design matrix with each column containing spread data for significant independent variables. Equation (1) can be re-written for each distribution parameter as:

$$g_1(\boldsymbol{\mu}^{(s)}) = \boldsymbol{\eta}_1^{(s)} = \mathbf{X}_1^{(s)}\boldsymbol{\beta}_1^{(s)} \tag{2}$$

$$g_2(\boldsymbol{\sigma}^{(s)}) = \boldsymbol{\eta}_2^{(s)} = \mathbf{X}_2^{(s)}\boldsymbol{\beta}_2^{(s)} \tag{3}$$

$$g_3(\boldsymbol{\nu}^{(s)}) = \boldsymbol{\eta}_3^{(s)} = \mathbf{X}_3^{(s)}\boldsymbol{\beta}_3^{(s)} \tag{4}$$

$$g_4(\boldsymbol{\tau}^{(s)}) = \boldsymbol{\eta}_4^{(s)} = \mathbf{X}_4^{(s)}\boldsymbol{\beta}_4^{(s)} \tag{5}$$

The choice of the link function influences how the linear predictor (i.e., systematic component $\mathbf{X}_k^{(s)} \boldsymbol{\beta}_k^{(s)}$) relates to each parameter. For example, a log link function for the standard deviation $g_2(\sigma^{(s)}) = log(\sigma^{(s)})$ results in the relationship $log(\sigma^{(s)}) = \boldsymbol{\eta}_2^{(s)} = \mathbf{X}_2^{(s)} \boldsymbol{\beta}_2^{(s)}$, hence the distribution parameter itself is obtained through a transformation $\sigma^{(s)} = \exp(\mathbf{X}_2^{(s)} \boldsymbol{\beta}_2^{(s)})$. Parameters $\boldsymbol{\theta}_k^{(s)}$ are calculated by maximizing the penalized likelihood using an RS algorithm which does not require calculation of the likelihood function's cross derivatives, but is a generalisation of the algorithm for fitting Mean and Dispersion Additive Models, (MADAM) as in [11]. The RS algorithm is a simple and fast method for larger data sets which does not require accurate starting values for distribution parameters to ensure convergence (constant default starting values are typically adequate) [12]. The RS algorithm is more suited for cases where the parameters of population probability density functions are information orthogonal and is reported by authors to be successfully used for the convergence of all the suitable distributions defined within the GAMLSS framework, although it may occasionally be slow.

To model electricity spreads within the GAMLSS approach, the distribution parameters sigma and tau should take positive values only but the mu and nu should be able to take either positive or negative values. Therefore distributions with the following link functions for each parameter are considered (see Table 1): location (mu): `identity`; scale (sigma): `log`; shape (nu): `identity`; shape (tau): `log`.

There are seven candidate distributions which we consider to be sufficiently flexible and which correspond to the necessary link function specifications: Johnson's SU, Johnson's original SU (JSU), skew power exponential type 1 (SEP1), skew power exponential type 2 (SEP2), skew t type 1 (ST1), skew t type 2 (ST2), and skew t type 5 (ST5). The Box–Cox power exponential and Box–Cox t distributions are not suitable due to only being defined on the positive interval $Y_t^{(s)} \in [0, +\infty)$, while the electricity spread data is highly skewed and kurtotic (see Figures 4 and 5). The skew t type 3 and 4 are also unsuitable since they are only defined for positive skewness yet the spread data is often negatively skewed.

**Table 1.** Continuous four parameter distributions—suitable distributions in the Generalised Additive Model for Location, Scale and Shape (GAMLSS).

| Continuous Distribution | $\mu$ | $\sigma$ | $\nu$ | $\tau$ |
|---|---|---|---|---|
| Johnson's SU (JSU) | identity | log | identity | log |
| Johnson's original SU (JSUo) | identity | log | identity | log |
| Skew power exponential type 1 (SEP1) | identity | log | identity | log |
| Skew power exponential type 2 (SEP2) | identity | log | identity | log |
| Skew t type 1 (ST1) | identity | log | identity | log |
| Skew t type 2 (ST2) | identity | log | identity | log |
| Skew t type 5 (ST5) | identity | log | identity | log |

The expected value of a random variable $Y_t$ for each distribution in Table 1 does not in general equal to the location parameter $\mu_t$. To be precise, it is given by $E(Y_t) = \mu_t + \sigma_t E(Z_t)$, where $z_t = \frac{y_t - \mu_t}{\sigma_t}$ is the standardized value of $y_t$, the expectation of $Z_t$ is not always zero and is specified for each distribution separately (see Appendix Equations (A1)–(A4) for details), and $\mu_t, \sigma_t, \nu_t, \tau_t$ are the location, scale and shape parameters of the given distribution at time step $t$. In the case of Johnson's SU and skew t type 1 distributions, the expected value does equal the distribution location parameter, $E(Y_t) = \mu_t$. Large fitted/forecasted tau values are capped at 100 for algorithmic convergence.

## 2.3. Distribution Selection

This section performs a thorough analysis of candidate distributions, shown in Table 1, to establish which of the distributions fits each spread hour data the best. This allows us to subsequently perform out-of-sample forecasting analysis using a rolling window estimation technique, presented in Section 3, where the best-chosen distribution is used to fit a model to each spread number and perform 1-step ahead forecasts on a rolling basis.

For model specifications, we begin by analysing which of the seven-candidate distributions fits each spread hour the best. We also analyse whether using a single distribution for modeling all of the spread data is a viable possibility as this might be useful for applications requiring analytical generality. The analysis is divided into two main steps: (a) simple distribution fit, where each distribution of Table 1 is fitted to the spreads in the training data set and the Akaike Information Criterion (AIC) is used to assess the goodness of fit; and (b) factor-based distribution fit, where for each spread hour, the candidate distributions resulting from the analysis of step (a) are used to build a model using exogenous factors. The fit of these models is assessed using a validation data set and a number of goodness of fit measures.

### 2.3.1. Methodology

To perform a thorough analysis when selecting the best distribution for each spread, we use a training data set, comprised of the first 60% of the spread time series. This is used for model fitting, following which a validation data set, comprised of the next 20% of the spread time series, is used for building forecasts and model evaluation. The validation set is useful when there is a question of model selection prior to final out-of-sample forecasting in the backtesting set of data [13]. The validation thus ensures that the predictive power of our trained models is evaluated on the unseen (validation) data set rather than on the fitting in-sample data. This is to avoid over-fitting and ensures that the final backtesting 20% of the data has neither been used for model fitting nor model selection.

The simple distribution fit analysis is performed using the training data, $\mathbf{Y}_{train} \in R^{1150 \times 276}$. Therefore, for each spread number $s = 1, ..., 276$ and each distribution $i = 1, ..., 7$ of Table 1, we build a model $\widehat{M}^{(s)} \leftarrow \mathbf{Y}_{train}^{(s)} \sim \mathbf{1}$, resulting in $\widehat{\mathbf{M}} \in \mathbb{R}^{276 \times 7}$ models. The AIC criteria of models $\widehat{\mathbf{M}}^{(s)} \in \mathbb{R}^7$, obtained for spread number $s$, are ranked in ascending order and the distribution corresponding to the model with the lowest AIC criterion is selected as the "distribution of best fit". Occasionally GAMLSS failed to fit the distribution to the spreads data, in which case that distribution was omitted for that spread. This problem was encountered for spread hours and distributions: 00–01 SEP1, 03–04 SEP2, 12–13 SEP1.

The results show that skew t type 5 distribution is selected most often as the distribution of best fit based on this simple selection criterion (see Table 2). It is closely followed by the same family Skew t type 1 distribution. Overall, all of the distributions except JSUo are indicated as potentially of the best fit for the training spread data. A more detailed breakdown of the spread hours for which each distribution was selected is displayed in Figure 6. We use the result to further analyse the six possible distributions with an factor-based distribution fit method.

**Table 2.** Simple distribution fit (training data)—best distribution selection based on AIC.

| JSU | JSUo | SEP1 | SEP2 | ST1 | ST2 | ST5 |
|-----|------|------|------|-----|-----|-----|
| 49 | 0 | 38 | 14 | 68 | 33 | 74 |

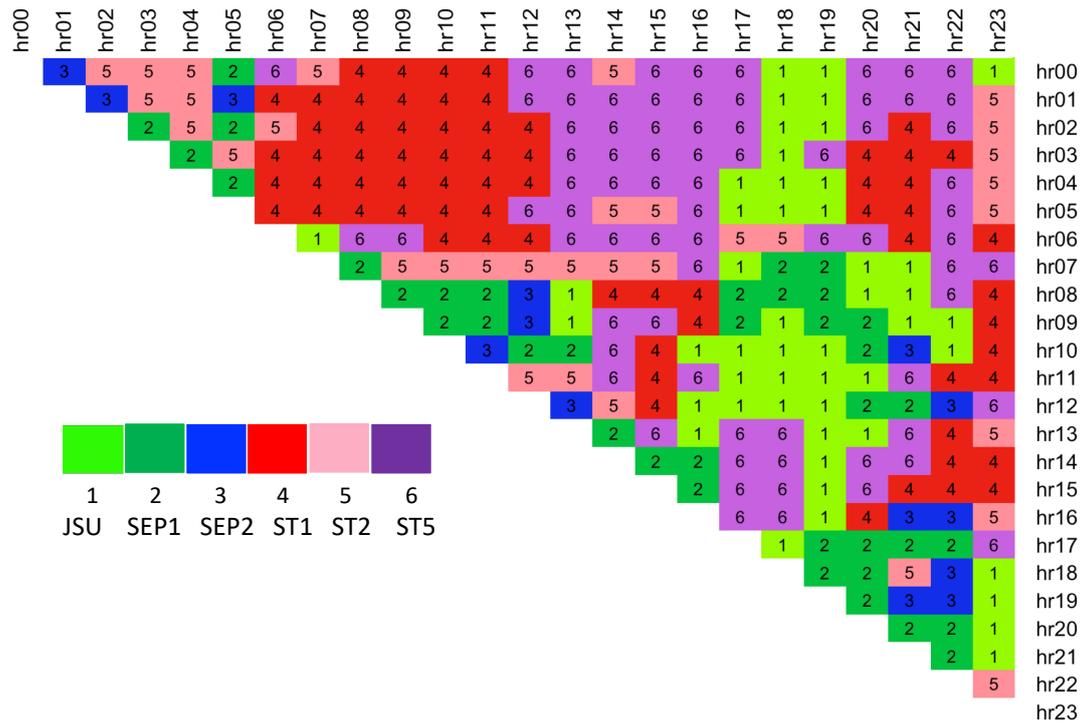

**Figure 6.** Continuous four-parameter distribution chosen as "best fit" to training spread data using `gamlss <-y ~ 1` and the AIC criterion.

The results of simple distribution fit indicated that six out of seven continuous four-parameter distributions could be used for modeling electricity spread data (see Table 2), which are $D^{(1)}$—JSU, $D^{(2)}$—SEP1, $D^{(3)}$—SEP2, $D^{(4)}$—ST1, $D^{(5)}$—ST2, and $D^{(6)}$—ST5. Hence we proceed with a factor-based analysis by estimating each time-varying parameters of the candidate distributions using exogenous variables within the GAMLSS framework. The training data set is comprised of the dependent $\mathbf{Y}_{train} \in R^{1150 \times 276}$ and independent $\mathbf{X}_{train} \in R^{1150 \times 9 \times 276}$ variables, where for each spread number $s$, under distribution $D^{(i)}, i = 1, ..., 6$, the initial equations relating each distribution parameter to its linear predictor are:

$$\hat{\mu}_t = \hat{\beta}_{1,0} + \hat{\beta}_{1,1}x_t + ... + \hat{\beta}_{1,8}x_{8,t} = \mathbf{x}_t^T \hat{\boldsymbol{\beta}}_1 \tag{6}$$

$$log(\hat{\sigma}_t) = \hat{\beta}_{2,0} + \hat{\beta}_{2,1}x_t + ... + \hat{\beta}_{2,8}x_{8,t} = \mathbf{x}_t^T \hat{\boldsymbol{\beta}}_2 \tag{7}$$

$$\hat{v}_t = \hat{\beta}_{3,0} + \hat{\beta}_{3,1}x_t + ... + \hat{\beta}_{3,8}x_{8,t} = \mathbf{x}_t^T \hat{\boldsymbol{\beta}}_3 \tag{8}$$

$$log(\hat{\tau}_t) = \hat{\beta}_{4,0} + \hat{\beta}_{4,1}x_t + ... + \hat{\beta}_{4,8}x_{8,t} = \mathbf{x}_t^T \hat{\boldsymbol{\beta}}_4 \tag{9}$$

where $\hat{\boldsymbol{\beta}}_k = [\hat{\beta}_{k,0}, \hat{\beta}_{k,1}, ..., \hat{\beta}_{k,8}]^T \in \mathbb{R}^9$ is initial vector of coefficients for distribution parameter $k$, $\mathbf{x}_t = [1, x_{1,t}, ..., x_{8,t}]^T \in \mathbb{R}^9$ is the initial vector of independent variables where $x_1$ is the spread of lagged day-ahead electricity price, $x_2$ is the gas Gaspool forward daily price, $x_3$ is the coal ARA forward daily price, $x_4$ is the spread of wind day-ahead forecast, $x_5$ is the spread of solar day-ahead forecast, $x_6$ is the dummy variable taking value 1 for weekends/holidays, $x_7$ is the spread of the day-ahead total load forecast, and $x_8$ is the interaction load variable.

The models are specified using iterative updating of the equations for each distribution parameter, where the refinement is performed by deleting the most insignificant variable one-by-one and re-estimating the model, until all variables are significant at 5% (see Algorithm 1). This results in $\hat{\mathbf{M}} \in \mathbb{R}^{276 \times 6}$ models containing the estimated coefficients. However during the process it became apparent that some distributions were not suitable for modeling certain spreads within the GAMLSS framework. Convergence was not achieved for some distribution parameters, typically $\tau$ and occasionally $\mu$ (see Appendix C.1) and whenever this happened the distribution was omitted from candidacy for that spread.

Evaluation

The models estimated using the factor-based distribution fit are analysed in four ways: (1) producing the expected value fit over the training data, (2) producing the expected value fit over validation data (comprising next 20% of unseen time series, at data points $t = 1151, ..., 1534$), (3) analysing the goodness of fit over the validation data using Root Mean Squared Error, and (4) analysing the goodness of fit over validation data using Pinball Loss function measure.

---

**Algorithm 1** Factor-based distribution fit—model specification and estimation

---

1: **for** each spread number $s = 1, ..., 276$ **do**
2:     Extract full training data design matrix $\mathbf{X}_{train}^{(s)} \in \mathbb{R}^{1150 \times 9}$ i.e., time steps $t = 1, ..., 1150$
3:     **for** each distribution $D^{(i)}, i = 1, ..., 6$ **do**
4:        Initialise iteration number $j \leftarrow 0$
5:        Initialise model $\widehat{M}_j^{(s,i)} \leftarrow \{\widehat{\boldsymbol{\beta}}_{k,j}^{(s,i)}\}_{k=1}^4 \in \mathbb{R}^{J_k+1 \times 4}$ using RS algorithm [11]
6:        **while** any $\widehat{\boldsymbol{\beta}}_{k,j}^{(s,i)}$ insignificant at 5% **do**
7:           $j \leftarrow j + 1$
8:           Find most insignificant coeff. $\widehat{\beta}^{\dagger}$ (except intercept) across all $k$
9:           Remove exogenous variable associated with $\widehat{\beta}^{\dagger}$ from Equation for $k$ and from $\mathbf{X}_{train}^{(s)}$
10:          Re-estimate model $\widehat{M}_j^{(s,i)} \leftarrow \{\widehat{\boldsymbol{\beta}}_{k,j}^{(s,i)}\}_{k=1}^4$ using RS algorithm

---

Expected Value Fit Over Training Data

Although the main focus of the methodology is to fit and forecast the density functions, it is a useful specification check to look at the calibration of the mean values. Therefore, the fitted distribution parameters $\widehat{\theta}_{train}^{(i)} \in \mathbb{R}^{1150 \times 4 \times 276}$ for each distribution $i = 1, ..., 6$ are used to find the fitted expected value $E(\mathbf{Y}_{train})$ of the training spread price data for $s = 1, ..., 276$ and training data points $t = 1, ..., 1150$ using expected value expression relevant to each distribution (see Appendix B).

Expected Value Fit Over Validation Data

The fitted models $\widehat{\mathbf{M}} \in \mathbb{R}^{276 \times 6}$, containing the estimated coefficients for each spread number $s = 1, ..., 276$ under each distribution $D^{(i)}, i = 1, ..., 6$, are used to forecast the distribution parameters, $\widehat{\theta}_{validate}^{(i)} \in \mathbb{R}^{383 \times 4 \times 276}$ over validation data. We note that the same estimated model $\widehat{M}^{(s,i)}$ is used to build the forecasts over validation time series, i.e., each vector of estimated coefficients $\widehat{\boldsymbol{\beta}}_k^{(s,i)}$ for distribution parameters $k = 1, ..., 4$ is re-used at each time step $t$ to make the predictions. Once the distribution parameters are forecasted the expected value of the spread is calculated using Appendix Equations (A1)–(A4).

Goodness of Fit Measure—Root Mean Squared Error

The Root Mean Squared Error (RMSE), was additionally used to check the accuracy of the forecasted expected value of the spreads over the validation data set. The RMSE is calculated for each spread $s = 1, ..., 276$ and for each distribution $D^{(i)}, i = 1, ..., 6$, using:

$$\text{RMSE}^{(s,i)} = \sqrt{\sum_{t=1}^{383} \left[ y_t^{(s,i)} - E\left(Y_t^{(s,i)}\right) \right]^2} \tag{10}$$

The distribution resulting in the lowest RMSE was selected as the best. As the expected value at each time step $t$ is calculated from all of the four forecasted parameters $(\widehat{\mu}_t^{(s,i)}, \widehat{\sigma}_t^{(s,i)}, \widehat{\nu}_t^{(s,i)}, \widehat{\tau}_t^{(s,i)})$, the RMSE measure provides a goodness of fit based on the forecast for the entire distribution specification.

The results are summarised in Table 3, which shows that ST5 distribution was used to form models corresponding to the lowest error across $\frac{110}{276} * 100 = 40\%$ of the spreads. This is in line with the finding of the simple distribution fit over the spread training data which indicated that ST5 is chosen as the best distribution most often (see Table 2). A detailed breakdown of best distribution assignment for each spread hour and the corresponding RMSE values are shown in Figures 7 and 8 respectively, where blue indicates a lower error. The results show that spreads with hours 13.00, 14.00 are the hardest to forecast since they have the highest RMSE error (dark red).

**Table 3.** Factor-based distribution fit (validation data)—best distribution based on  root mean squared error (RMSE).

| JSU | SEP1 | SEP2 | ST1 | ST2 | ST5 |
|-----|------|------|-----|-----|-----|
| 49  | 31   | 21   | 11  | 60  | 110 |

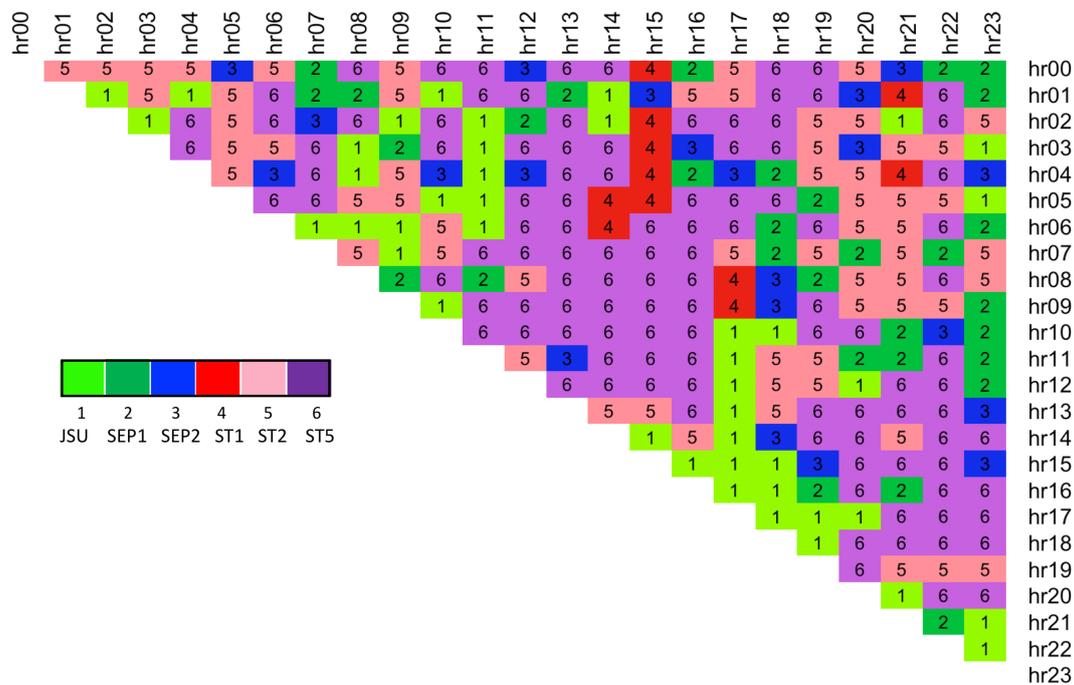

**Figure 7.** Best distributions based on lowest RMSE when forecasting expected value of spread price, $E(Y)$, over validation horizon.

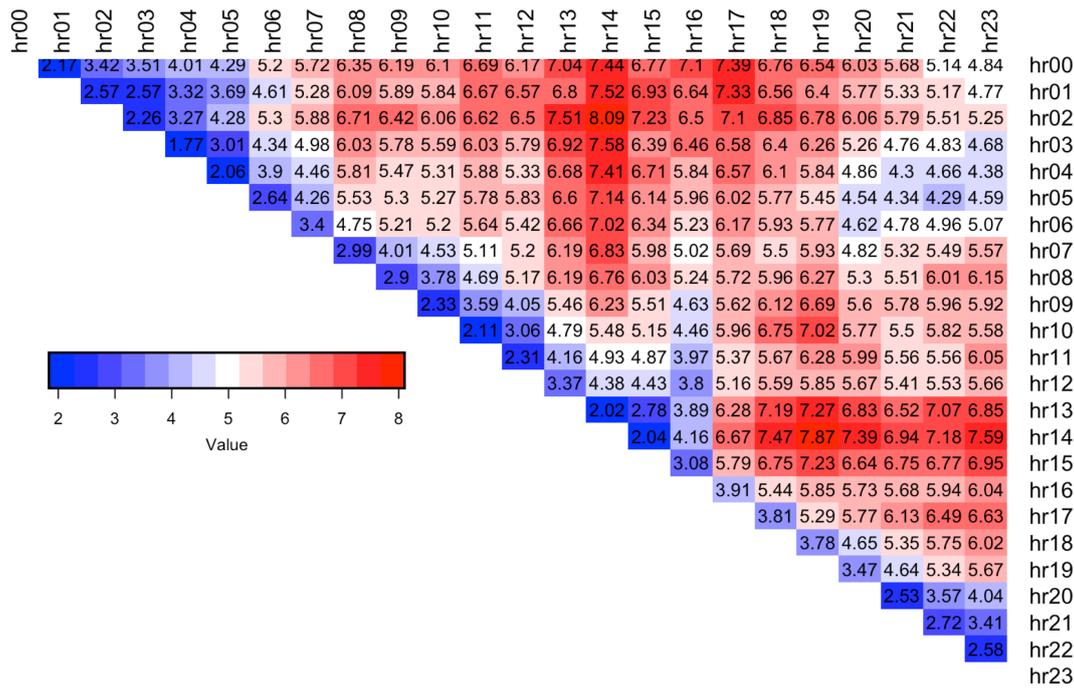

**Figure 8.** RMSE values corresponding to selected best distributions when forecasting expected value of spread price, $E(Y)$, over validation horizon.

### Goodness of Fit Measure—Pinball Loss Score

As our main modeling focus is upon estimating the entire four-parameter distributions fitted to the spread data at each time step, we focus next upon the calibration of the estimated quantiles, using the Pinball Loss (PL) function [14]. The PL function is always positive, and has a higher value the further away the quantile estimate is away from the target value. The quantiles are deemed to be fitted the best when the Pinball Loss is at its lowest. The quality of quantile fit is important for Value-at-Risk compliance in the context of risk management applications. We adopt this measure as our main performance metric to select the best distribution for each spread. We follow Algorithm 2 when making an assessment of each predictive density power and outline the steps involved in calculating the performance measure below.

1.  Pinball Loss Values. At each time step $t$, the target quantiles $q_a, a = 1, 2, ..., 99$ are extracted by inverting the cumulative distribution function of distribution $D^{(i)}, i = 1, ..., 6$ specified with forecasted distribution parameters $(\hat{\mu}_t^{(s,i)}, \hat{\sigma}_t^{(s,i)}, \hat{\nu}_t^{(s,i)}, \hat{\tau}_t^{(s,i)})$ and compared to the realisation of true spread price, $y_t$, using:

$$L_t^{(s,i)}(q_a, y_t) = \begin{cases} \frac{1-a/100}{q_{a,t}-y_t} & \text{if } y_t < q_a \\ \frac{a/100}{y_t - q_{a,t}} & \text{if } y_t \geq q_a \end{cases} \tag{11}$$

This results in Pinball Loss Values vector, $\mathbf{L}_t^{(s,i)}(\mathbf{q}, y_t) \in \mathbb{R}^{J_a}$, where $J_a = \{99, 97, 95\}$ is the number of quantiles extracted. Certain quantiles failed to converge when calculating the full set of 99 values due to convergence issues around distribution tails. To resolve this we removed the tail quantiles one pair at a time and attempted to extract the quantiles again (i.e., resulting in 97 quantiles: $[q_2, q_{98}]$ and if still not convergent 95 quantiles $[q_3, q_{97}]$). If this still did not resolve the issue, we omitted calculating the Pinball Loss value for that time step and recorded the number of such occurrences, $n$. Note: JSU and ST5 extracted quantiles 100% of the time, making these two distributions the most stable out of the 6 tested. This is in contrast to ST1 and ST2 distributions which failed to converge often (see Appendix C Table A1 for the number of times each distribution failed to extract a quantile).

2. Pinball Loss Scores. At each time step, $t$, the average over Pinball Loss Values is obtained resulting in a single Pinball Loss Score, $\bar{L}_t^{(s,i)}(q_a, y_t)$:

$$\bar{L}_t^{(s,i)}(q_a, y_t) = \sum_{a=1}^{J_a} L_t^{(s,i)}(q_a, y_t) \tag{12}$$

This results in a vector $\bar{\mathbf{L}}^{(s,i)}(\mathbf{q}, \mathbf{y}) \in \mathbb{R}^{J_b}$, where $J_b = 383 - n$ and $n$ denotes the number of occurrences when $L_t^{(s,i)}(q_a, y_t)$ was not available for calculation.

3. Pinball Loss Performance Measure. The final step in calculating a single value describing the goodness-of-fit measure using the Pinball Loss function is finding the average of the Pinball Loss Scores across the full validation forecast horizon at time steps $t = 1151, ..., 1534$ (i.e., 383 days):

$$\mathcal{L}^{(s,i)} = \sum_{t=1}^{J_b} \bar{L}_t^{(s,i)}(q_a, y_t) \tag{13}$$

The distribution $D^{(i)}$ corresponding to the model with the lowest Pinball Loss Performance Measure, $\mathcal{L}^{(s,i)} \in \mathbb{R}$, is selected as the distribution of best fit for spread number $s$.

---

**Algorithm 2** Model selection using Pinball Loss function

---

1:  **for** each spread number $s = 1 : 276$ **do**
2:　　Extract validation data design matrix $\mathbf{X}_{validate}^{(s)} \in \mathbb{R}^{383 \times 9}$ i.e., time steps $t = 1151, ..., 1534$
3:　　**for** each distribution $D^{(i)}, i = 1 : 6$ **do**
4:　　　　Forecast *parameters* of $D^{(i)}$ resulting in $\widehat{\boldsymbol{\theta}}^{(s,i)} = [\widehat{\boldsymbol{\mu}}^{(s,i)}, \widehat{\boldsymbol{\sigma}}^{(s,i)}, \widehat{\boldsymbol{\nu}}^{(s,i)}, \widehat{\boldsymbol{\tau}}^{(s,i)}]^T$ each $\in \mathbb{R}^{383}$
5:　　　　**for** time step $t = 1, ..., 383$ **do**
6:　　　　　　Obtain vector of quantiles $\mathbf{q}_t^{(s,i)} \in \mathbb{R}^{J_a}$ using $\widehat{\boldsymbol{\theta}}_t^{(s,i)}$
7:　　　　　　**for** each quantile $q_{a,t}$ **do**
8:　　　　　　　　Calculate Pinball Loss value $L_t^{(s,i)}(q_a, y_t)$ using Equation (11)
9:　　　　　　　　Calculate Pinball Loss Score $\bar{L}_t^{(s,i)}(q_a, y_t)$ using Equation (12)
10:　　　　Calculate Pinball Loss Performance Measure $\mathcal{L}^{(s,i)}$ using Equation (13)
11:　　Find $n = \text{argmin}_i \mathcal{L}^{(s)}$ and select corresponding distribution $D^{(n)}$ as best fit for $s$

---

The Pinball Loss Performance Measures are analysed for each spread $s$ and the results are presented in a $24 \times 24$ upper diagonal matrix, containing the best distribution number selected for each intraday spread between the hours indicated by the row and column labels (see Figure 9). The figure shows that the majority of the spreads are fitted best with ST5 distribution (number 6, purple colour). This is in line with the results obtained from the factor-based distribution fit using the RMSE measure (see Figure 7), and with the simple best distribution fit given by $y \sim 1$ function (see Figure 6), which both favoured the ST5 distribution.

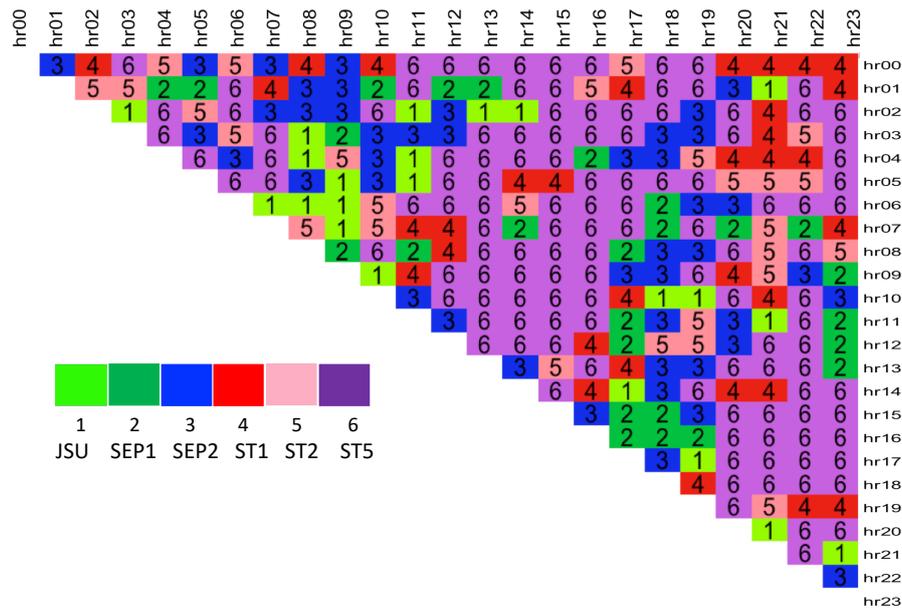

**Figure 9.** Best distribution based on the Pinball Loss function (validation data).

Table 4 specifies the number of times each distribution was selected as the best from the 276 possible spreads. While ST5 is selected for almost 50% of the spreads, the other 5 distributions were selected nearly with the same proportion for the remaining half of the spread data.

**Table 4.** Factor-based distribution fit (validation data)—best distribution based on Pinball Loss function.

| JSU | SEP1 | SEP2 | ST1 | ST2 | ST5 |
|-----|------|------|-----|-----|-----|
| 22  | 26   | 45   | 33  | 27  | 123 |

Further to this, we calculate the % difference of the best distribution PL score vs. that of the next best distribution, $\frac{|\mathcal{L}_1^{(s)} - \mathcal{L}_2^{(s)}|}{\mathcal{L}_2^{(s)}} * 100\%$, for each spread number $s$. When ST5 was selected as the best distribution, the average difference of the performance measure is 4.84% compared to, when other distributions were selected as best, the average difference is 1.44% (with approx 1/3 of these cases containing ST5 is the second-best). This makes the case for ST5 to potentially be used as a general best fit distribution across all spreads.

## 3. Rolling Window Forecast Analysis

We perform an out of sample forecasting analysis based on the best distribution selected for fitting each spread number using a rolling window estimation technique and compare the results to the performance of a Normal density as a benchmark. We first assess the performance of models fitted using the Pinball Loss function, and then we use the Diebold–Mariano test. The estimation horizon (rolling window size) comprises of 80% of the data ($T_0 = 1534$ observations), which is moved up 1 time step at a time.

For each spread we re-specify and re-estimate models using the chosen distribution for that spread (given by Figure 9) while following the rolling window approach, outlined in Figure 10. Starting at time step $t_1$ a model is specified and estimated over a fixed length horizon of $T_0$ observations (green coloured bracket). The equations for distribution parameters are obtained using iterative improvement, where the most insignificant variables are deleted one-by-one until all of the variables in each parameter's equation are significant at 5% level. A one-step ahead forecast is made and a full predictive density is obtained, $f_1$ (green colour). The window is moved by a 1-time step (blue colour) and the procedure is repeated until all of the forecasts are created for the current spread (comprising 383 data points). This procedure is applied to other spreads until all rolling window forecasts are obtained for 276 spreads (see Algorithm 3).

---

**Algorithm 3** Forecasting using rolling window

---

1: **for** each spread number $s = 1 : 276$ **do**
2:     Extract the best chosen distribution, $D^{(s)}$, for the current spread $s$, using Figure 9
3:     **for** forecast number $t = 1, ..., 383$ **do**
4:         Extract training data design matrix $\mathbf{X}_{train}^{(s,t)}$, of size $T_0 \times 9$, where $T_0 = [t, t + 1534 - 1]$
5:         Initialise iteration number $j \leftarrow 0$
6:         Initialise model $\widehat{M}_j^{(s,t)} \leftarrow \{\widehat{\boldsymbol{\beta}}_{k,j}^{(s,t)}\}_{k=1}^4 \in \mathbb{R}^{l_k+1 \times 4}$ using RS algo, for $k = 1, ..., 4$ parameters
7:         **while** any $\widehat{\beta}_{k,j}^{(s,t)}$ insignificant at 5% $\forall k$ **do**
8:             $j \leftarrow j + 1$
9:             Find most insignificant coeff. $\widehat{\beta}^\dagger$ (except intercept) across all $k$
10:            Remove exogenous variable associated with $\widehat{\beta}^\dagger$ from eq. for $k$ and from $\mathbf{X}_{train}^{(s,t)}$
11:            Re-estimate model $\widehat{M}_j^{(s,t)} \leftarrow \{\widehat{\boldsymbol{\beta}}_{k,j}^{(s,t)}\}_{k=1}^4$ using RS algorithm
12:        Using model $\widehat{M}^{(s,t)}$ create 1-step ahead forecast resulting in $\widehat{\boldsymbol{\theta}}_t^{(s)} = [\widehat{\mu}_t^{(s)}, \widehat{\sigma}_t^{(s)}, \widehat{v}_t^{(s)}, \widehat{\tau}_t^{(s)}]^T$
13:        Calculate forecasted expected value $E(Y_t)$ for time step $T_0 + 1$ using Equation for $D^s$ and $\widehat{\boldsymbol{\theta}}_t^{(s)}$

---

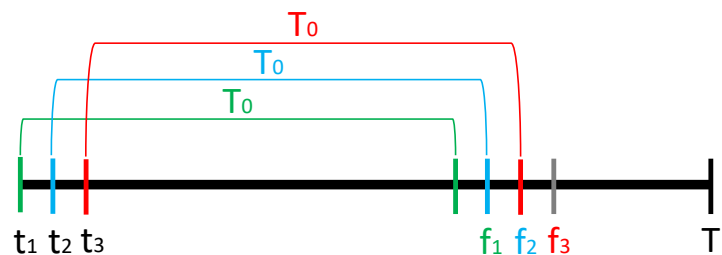

**Figure 10.** Rolling window scheme.

The resulting data structure is of the size: $276 \times 383$ and contains time-series of forecasts made for each spread. The rolling window forecast procedure was parallelised across 8 cores on an Intel Core i7 2.9 GHz processor and took 216 hours to complete. The same approach is used to form forecasts for the Normal Distribution resulting in a data structure containing 383 forecasts for each of the 276 electricity spreads.

### 3.1. In-Sample Evaluation

To validate the need for dynamic modeling, we display the evolution of fitted parameters for four example spreads obtained for the first rolling window training data (i.e., first 1534 data points used for specification and estimation phase), giving their associated skew type distributions: 00–08 (ST1), 08–12 (ST1), 12–16 (ST1), 16–20 (ST5).

#### 3.1.1. Explanatory Variable Coefficients

The variations in the size and sign of explanatory variable coefficients for different spreads of the same day illustrate the varying impact of drivers (see Tables 5 and 6 which show the estimated coefficients for selected 4 spreads: 00–08, 08–12, 12–16, 16–20). The coefficients were extracted from the model estimated for the first rolling window frame i.e., $t_1 - t_{1534}$, and obtained on standardized regressors, but without standardizing the dummy variable. The displayed values for the coefficients are all significant at 5% thus forming the equation for that *parameter*. Missing values indicate that the independent variable was not significant and thus was omitted from the equation specification for that *parameter* (column). The coefficients correspond to their associated independent variables: $\beta_1$ coeff for spread of the lagged day-ahead electricity price; $\beta_2$ coeff for gas Gaspool forward daily price; $\beta_3$ coeff for coal ARA forward daily price; $\beta_4$ coeff for spread of wind day-ahead forecast;

$\beta_5$ coeff for spread of solar day-ahead forecast; $\beta_6$ coeff for dummy variable taking value of 1 for weekends/holidays; $\beta_7$ coeff for spread of the day-ahead total load forecast; $\beta_8$ coeff for an interaction load variable.

**Table 5.** Illustration of coefficients for 4 distribution parameters of spreads 00–08 and 08–12.

|  | Spread: 00–08 | | | | Spread: 08–12 | | | |
|---|---|---|---|---|---|---|---|---|
|  | $\hat{\mu}$ | $log(\hat{\sigma})$ | $\hat{v}$ | $log(\hat{\tau})$ | $\hat{\mu}$ | $log(\hat{\sigma})$ | $\hat{v}$ | $log(\hat{\tau})$ |
| $\hat{\beta}_0$ | −16.12 | 1.91 | −1.17 | 2.11 | 3.68 | 1.83 | 1.37 | 1.71 |
| spr$_{t-1}$, $\hat{\beta}_1$ | 2.31 | −0.09 | −0.22 | 0.39 | 2.12 |  | −0.28 |  |
| gas, $\hat{\beta}_2$ | −2.62 | 0.22 |  | 0.56 | 0.73 | 0.22 |  |  |
| coal, $\hat{\beta}_3$ | 1.64 | 0.06 | −0.32 |  | −2.25 | 0.12 | 0.40 |  |
| wind spr, $\hat{\beta}_4$ | −1.49 |  | −0.46 | −0.63 | −1.94 |  |  |  |
| solar spr, $\hat{\beta}_5$ | −2.51 | 0.06 |  |  | −2.58 | −0.12 | −0.55 |  |
| dummy, $\hat{\beta}_6$ | 14.22 | −0.39 | 1.71 |  | −4.70 | −0.38 | −1.09 |  |
| load spr, $\hat{\beta}_7$ | 7.73 | 0.77 | −3.36 |  | −17.70 |  | 5.76 |  |
| int load, $\hat{\beta}_8$ | −8.02 | −0.48 | 3.57 |  | 17.59 |  | −5.39 |  |

**Table 6.** Illustration of coefficients for 4 distribution parameters of spreads 12–16 and 16–20.

|  | Spread: 12–16 | | | | Spread: 16–20 | | | |
|---|---|---|---|---|---|---|---|---|
|  | $\hat{\mu}$ | $log(\hat{\sigma})$ | $\hat{v}$ | $log(\hat{\tau})$ | $\hat{\mu}$ | $log(\hat{\sigma})$ | $\hat{v}$ | $log(\hat{\tau})$ |
| $\hat{\beta}_0$ | 1.48 | 1.31 | −0.11 | 2.11 | −4.31 | 1.42 | 0.001 | −1.73 |
| spr$_{t-1}$, $\hat{\beta}_1$ | 0.55 | −0.12 |  |  | 3.88 |  | −0.05 | 0.17 |
| gas, $\hat{\beta}_2$ | −1.41 |  | 0.41 | −0.60 |  | 0.22 | −0.03 |  |
| coal, $\hat{\beta}_3$ | 1.61 | 0.13 | −0.40 |  | 0.8 3 |  | −0.05 |  |
| wind spr, $\hat{\beta}_4$ | −2.46 |  |  | −0.30 | −1.45 |  | −0.04 |  |
| solar spr, $\hat{\beta}_5$ | −3.41 |  | −0.15 |  | −1.49 |  | −0.10 |  |
| dummy, $\hat{\beta}_6$ |  |  |  |  | −3.04 | 0.13 | −0.15 | 0.35 |
| load spr, $\hat{\beta}_7$ | 0.96 | −0.14 |  | −0.37 | −12.08 | 0.37 | 0.49 | −1.49 |
| int load, $\hat{\beta}_8$ |  |  | 0.39 |  | 11.47 | −0.40 | −0.39 | 1.36 |

Overall the signs and significances of the coefficients are intuitive. In particular, wind and solar production spreads have negative effects on the mu and nu parameters of spreads for the morning and afternoon spread pairs shown. Recall that the spread is defined as the former minus the later hours and so a higher wind and solar production spread will generally reduce the average spreads and also the skewness (since nu is related to skewness). This is consistent with the effects of wind and solar production on price levels reported in [9] and elsewhere.

### 3.1.2. Parameter Evolution—Year

The evolution of the four parameters throughout four years using 4 spread hours is examined i.e., establishing how selected spreads behave throughout each year. We selected four years: 2012, 2013, 2014, 2015 in order to depict the evolution and the changing dynamics of the parameters with time. The evolution of each distribution parameter is plotted on separate graphs (see Figure 11 for the evolution of $\hat{\mu}$, Figure 12 for evolution of $\hat{\sigma}$, Figure 13 for the evolution of $\hat{v}$, Figure 14 for evolution of $\hat{\tau}$). The results show that the fitted location parameter is in line with what would be expected for the realisation of true spread price, $y_t$, where the spreads between 08–12 hours tend to be positive (i.e., later hour is cheaper), while the 16–20 spreads are negative, i.e., later hour is more expensive. The standard deviation is highest for the spreads between night-time (less busy) and early morning/afternoon (green and blue lines). While the parameter nu tends to be positive for the 08–12 hours, and negative for the 00–08 hours.

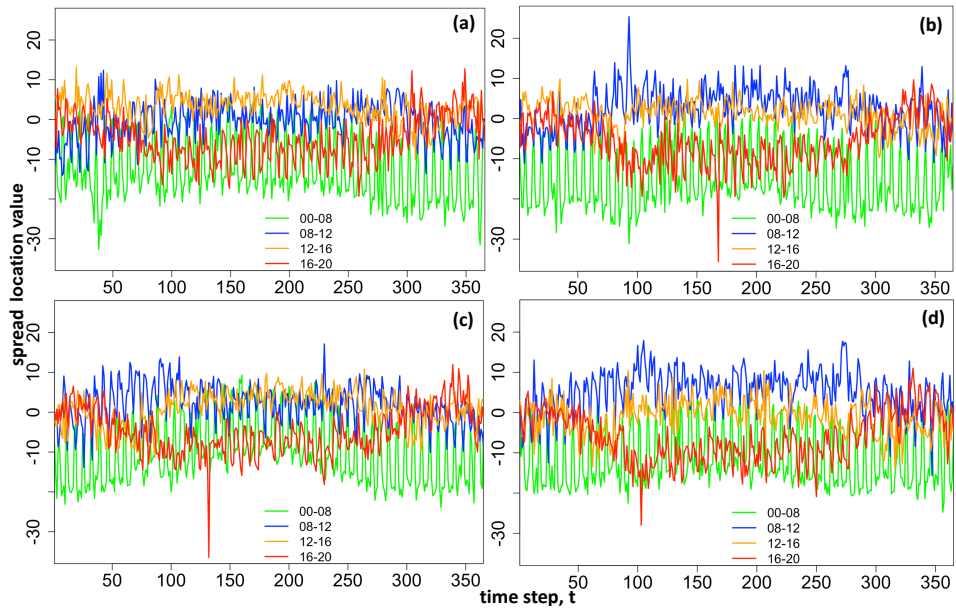

**Figure 11.** Evolution of fitted location parameter, $\hat{\mu}$, of the best distribution for years: (**a**–**d**) 2012–2015.

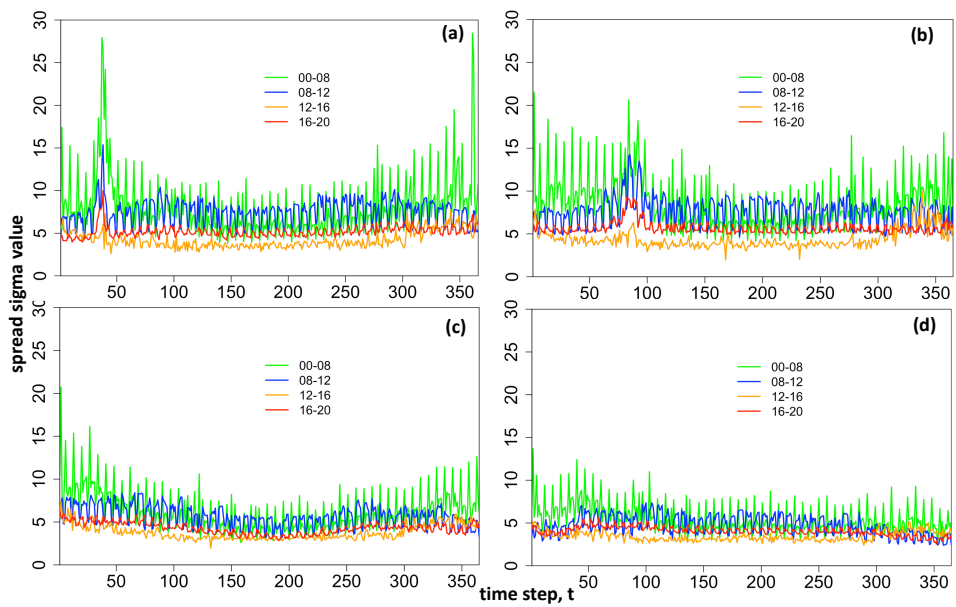

**Figure 12.** Evolution of fitted standard deviation of the best distribution, $\hat{\sigma}$, for years: (**a**–**d**) 2012–2015.

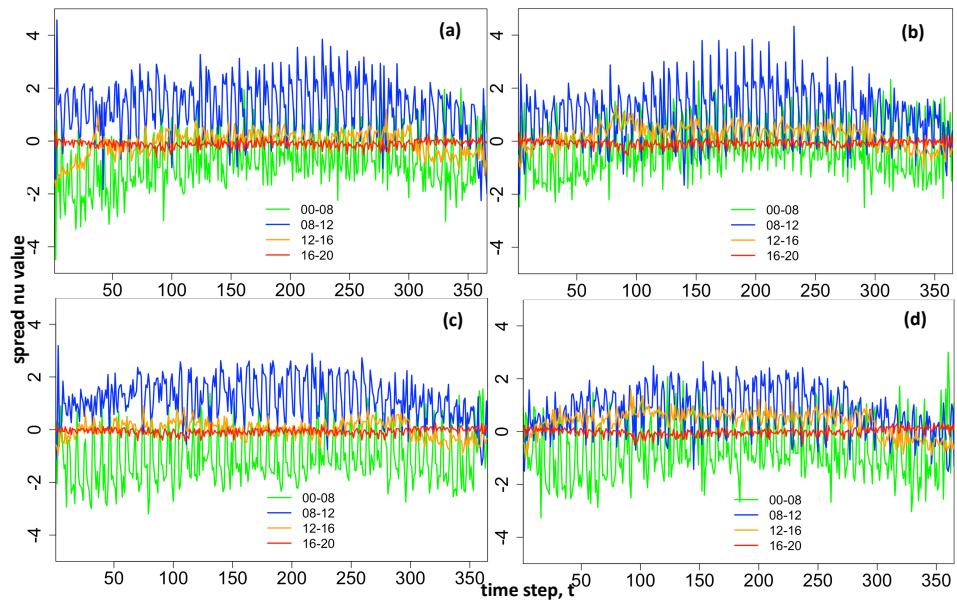

**Figure 13.** Evolution of fitted parameter nu of the best distribution, $\hat{\nu}$, for years: (**a–d**) 2012–2015.

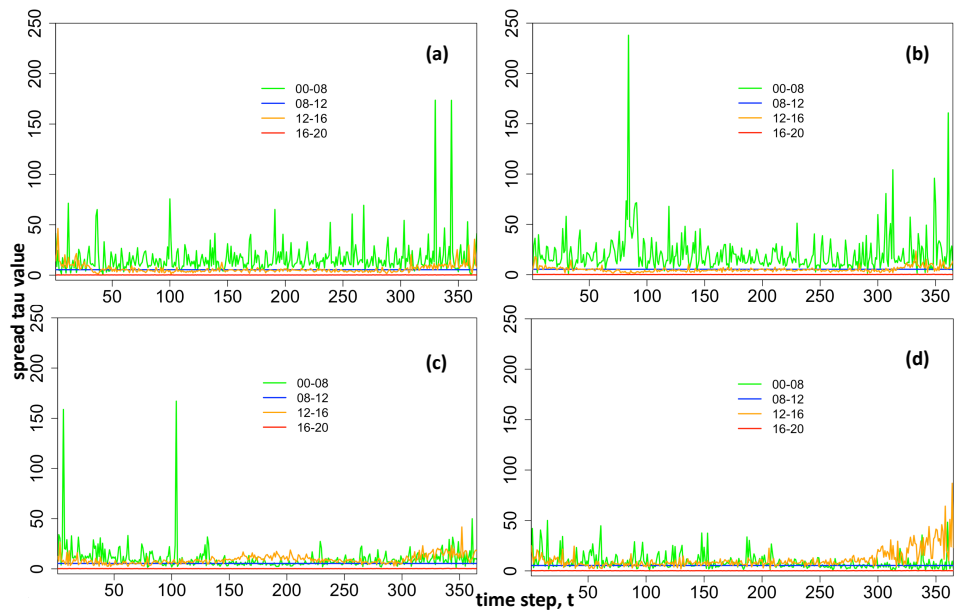

**Figure 14.** Evolution of fitted parameter tau of the best distribution, $\hat{\tau}$, for years: (**a–d**) 2012–2015.

Next, we plot realisations of spread prices, $y_t$, vs. expected value of fitted spread prices $E(Y_t)$ over the four years, picking different spread hours to show a variety of underlying skew type distributions: 00–09 (SEP2), 08–11 (SEP1), 11–19 (ST2), 16–22 (ST5). Figure 15 shows the evolution of realised spread prices, compared to the evolution of expected values produced by the estimated models (see Figure 16, note slightly smaller scale). The fitted expected values follow the realised spread prices pattern throughout each year, for example, the 16–22 spread tends to have more negative values in the summer time (i.e., electricity at earlier hour is less expensive) and positive values in the winter time (i.e., electricity at earlier hour is more expensive). It can be seen that the fitted spread values are slightly under-fitted as indicated by the difference in plot scale.

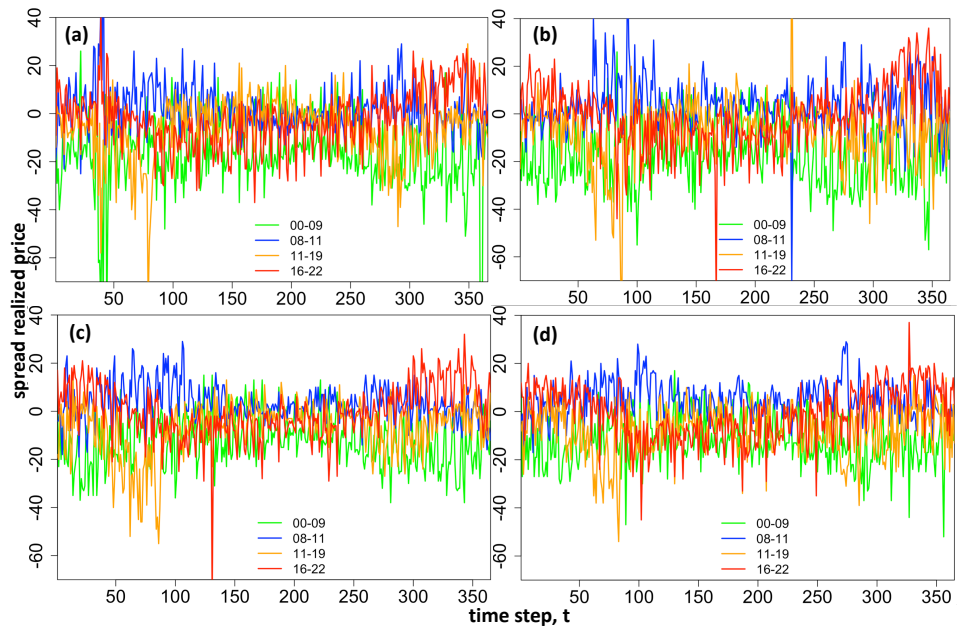

**Figure 15.** Evolution of realised spread prices, $y_t$, for years: (**a**–**d**) 2012–2015.

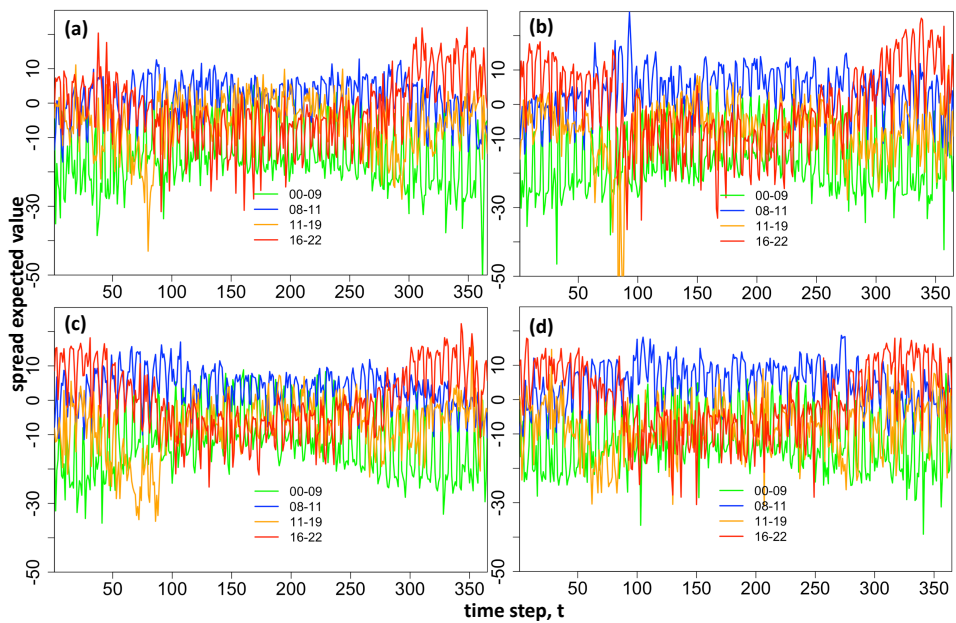

**Figure 16.** Evolution of fitted spread expected value, $E(Y)$, for years: (**a**–**d**) 2012–2015.

### 3.1.3. Parameter Evolution—Day

Figure 17 demonstrates the changing dynamics of the parameters throughout a day as well as seasons of the year, where subplots show evolution of (a) mu, (b) sigma, (c) nu and (d) tau. Each parameter is plotted using data for all 276 spread prices on the 1st of month of January (green line), March (blue line), June (orange line) and September (red line) of 2015. The plots thus depict the evolution of the four parameters throughout a day and throughout different times of the year.

The spreads were plotted sequentially starting with 23 spreads for midnight hour with all other hours of the day, i.e., 00–01, 00–02,...,00–23 (see grey dotted box of Figure 17a), followed by 22 spreads of hour 01 with all other hours and continued until the last spread of hour 22 with 23, therefore resulting in all 276 spreads being plotted on a single line. The results show periodic spikes due to points where the spread moves between 00 with all others, 01 with all others, etc. The overall plot also captures the changing dynamics of the parameters throughout seasons of the year.

Next we focus on the evolution of the fitted parameter mu for four chosen hours (midnight, 8am, noon, 4pm). Figure 18a zooms in on the dotted grey section of Figure 17a and depicts spreads between the midnight hour with all hours of the day. This zoomed in plot thus allows a closer examination of the evolution of the fitted parameter mu during a day for the 4 chosen days, and highlights the different behaviour not only within the day but during different season's of the year 2015. Figure 18b–d provide similar information for the spreads between hour: (b) 08 (c) noon (d) 16 with all other hours of the day (note the x-axis range decreases since there are less hours between which the spreads are calculated).

Likewise Figures 19–21 provide the plots for selected hours spread with all hours that follow it in that day for volatility, nu and tau respectively.

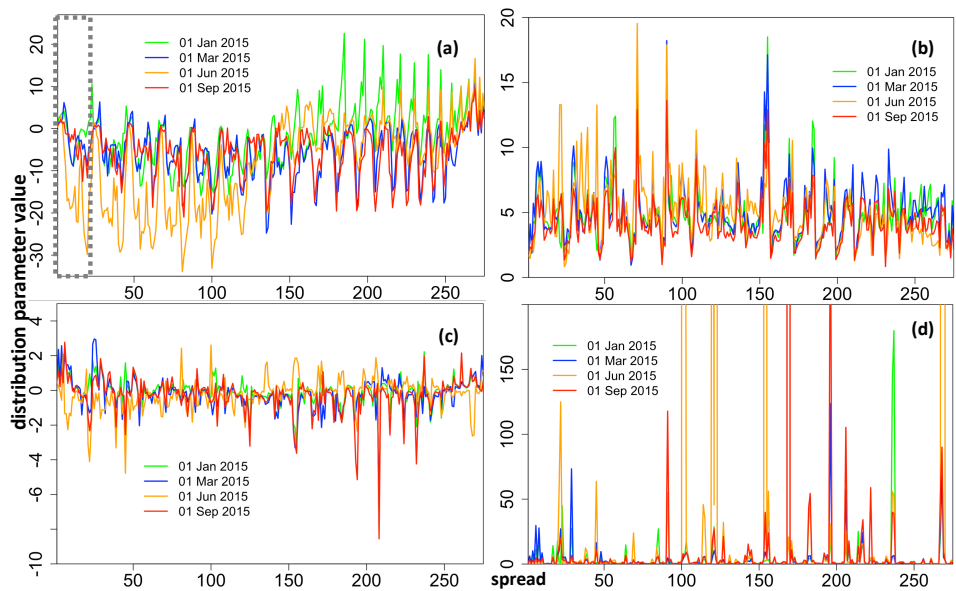

**Figure 17.** Evolution of fitted distribution parameters throughout the day depicted for 4 days of year 2015: (**a**) mu (grey box shows midnight spread with all other hours); (**b**) sigma (**c**) nu and (**d**) tau.

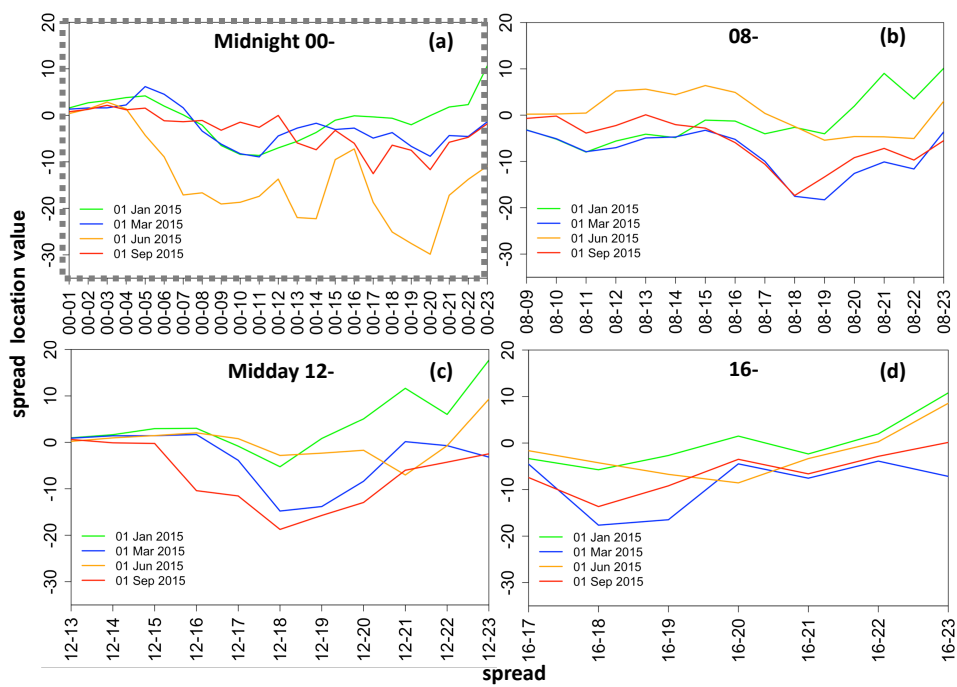

**Figure 18.** Evolution of fitted $\hat{\mu}$ throughout the day plotted for 4 chosen hours for the year 2015. Plots show hours spread with all other hours of the day that follow it: (**a**) 00; (**b**) 08; (**c**) 12 and (**d**) 16.

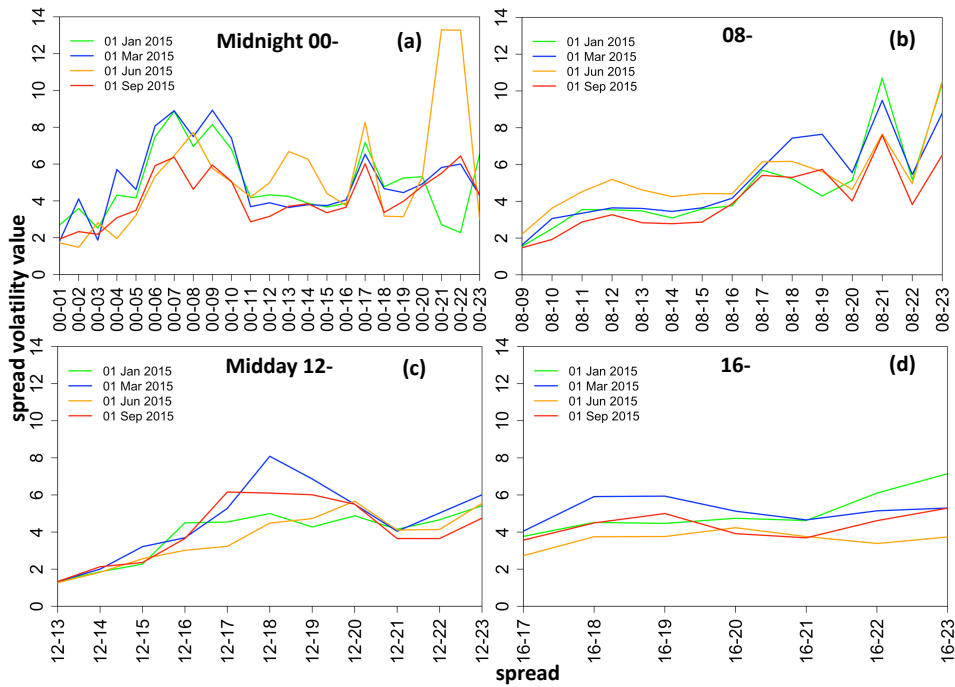

**Figure 19.** Evolution of fitted $\hat{\sigma}$ throughout the day plotted for 4 chosen hours for the year 2015. Plots show hours spread with all other hours of the day that follow it: (**a**) midnight, (**b**) 08, (**c**) 12 and (**d**) 16.

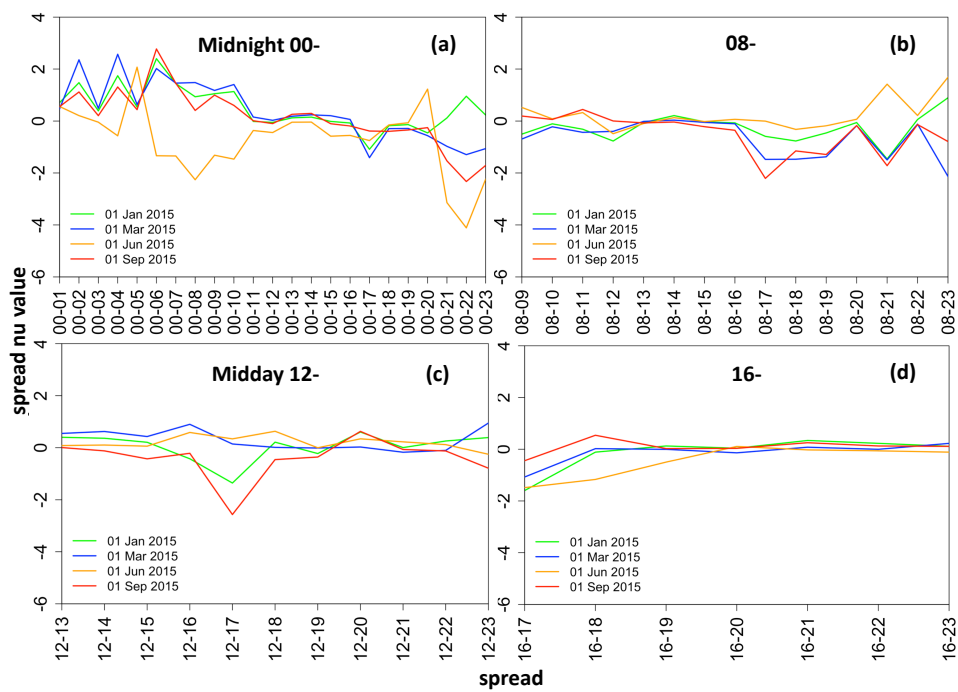

**Figure 20.** Evolution of fitted $\hat{\nu}$ throughout the day plotted for 4 chosen hours for the year 2015. Plots show hours spread with all other hours of the day that follow it: (**a**) midnight; (**b**) 08; (**c**) 12 and (**d**) 16.

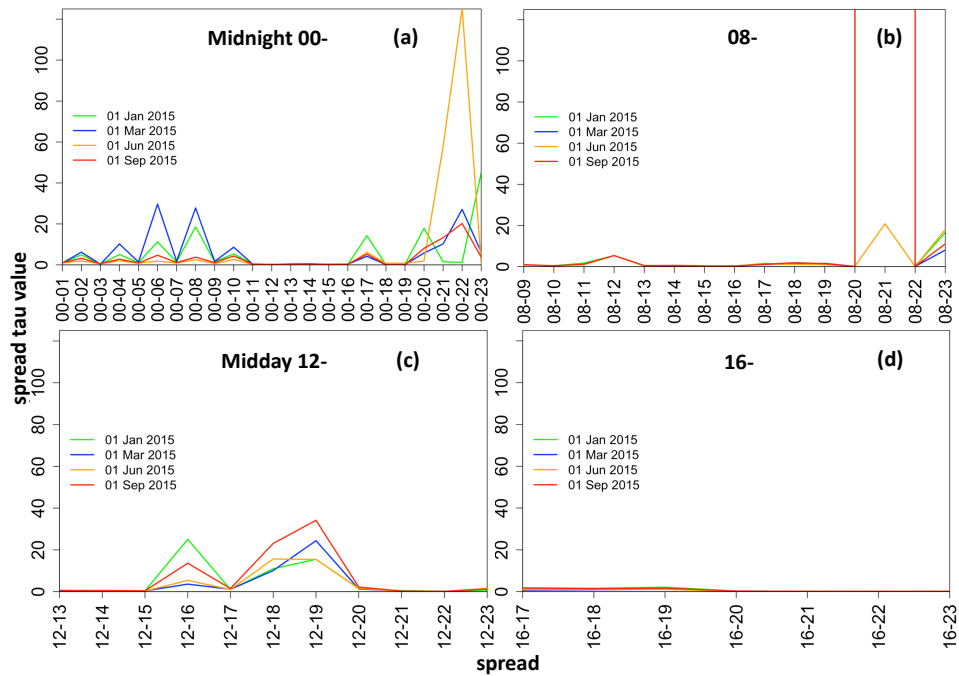

**Figure 21.** Evolution of fitted $\hat{\tau}$ throughout the day plotted for 4 chosen hours for the year 2015. Plots show hours spread with all other hours of the day that follow it: (**a**) midnight; (**b**) 08; (**c**) 12 and (**d**) 16.

### 3.2. Out-Of-Sample Evaluation

#### 3.2.1. RMSE-Based Evaluation

The first stage of out-of-sample evaluation of the obtained forecasts involves calculating the Root Mean Squared Error for skew type distributions and the Normal distribution. The RMSE values of skew type distribution models for the forecasted expected value of the rolling window test data are given in Figure 22. The results show that night-time spreads with hours 14.00, 17.00 and 18.00 had the biggest errors when forecasting the level. The RMSE values of the benchmark Normal type models for the forecasted expected value of the rolling window test data are given in Figure 23. The results show that night-time spreads with hours 14.00, 17.00 and 18.00 also had the biggest errors when forecasting the level. In order to thoroughly compare the performance of skew type distributions against the Normal benchmark, formal statistical testing is carried out next using Pinball Loss based measure.

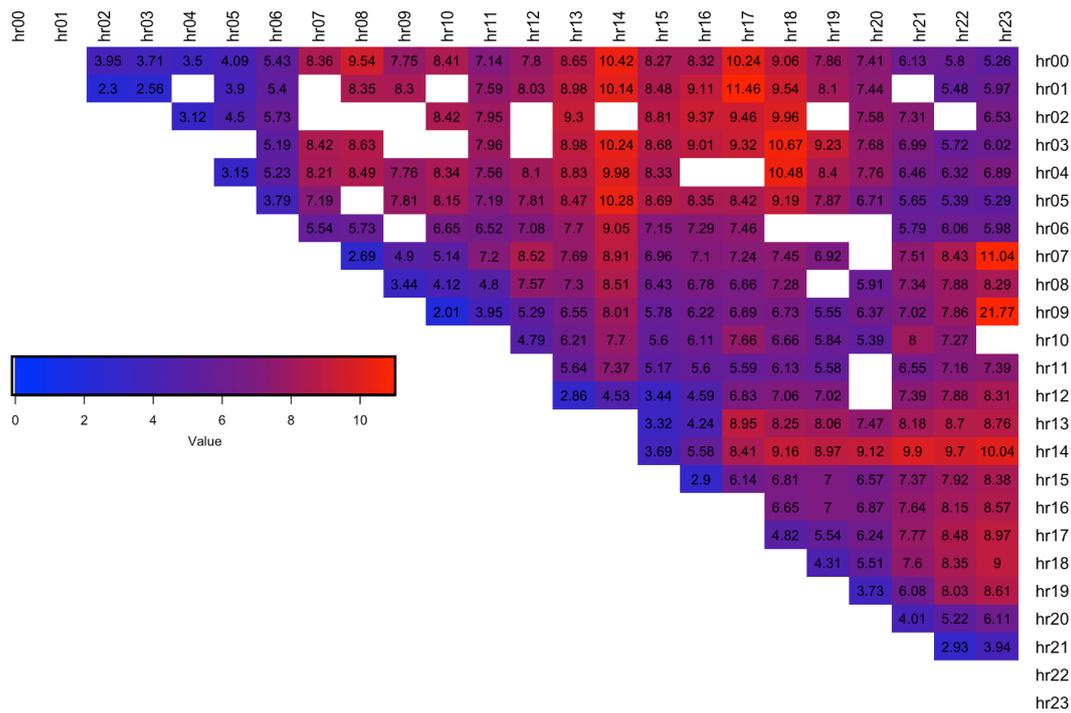

**Figure 22.** RMSE of skew type distributions for forecasting expected value, $E(Y)$, over rolling window test data.

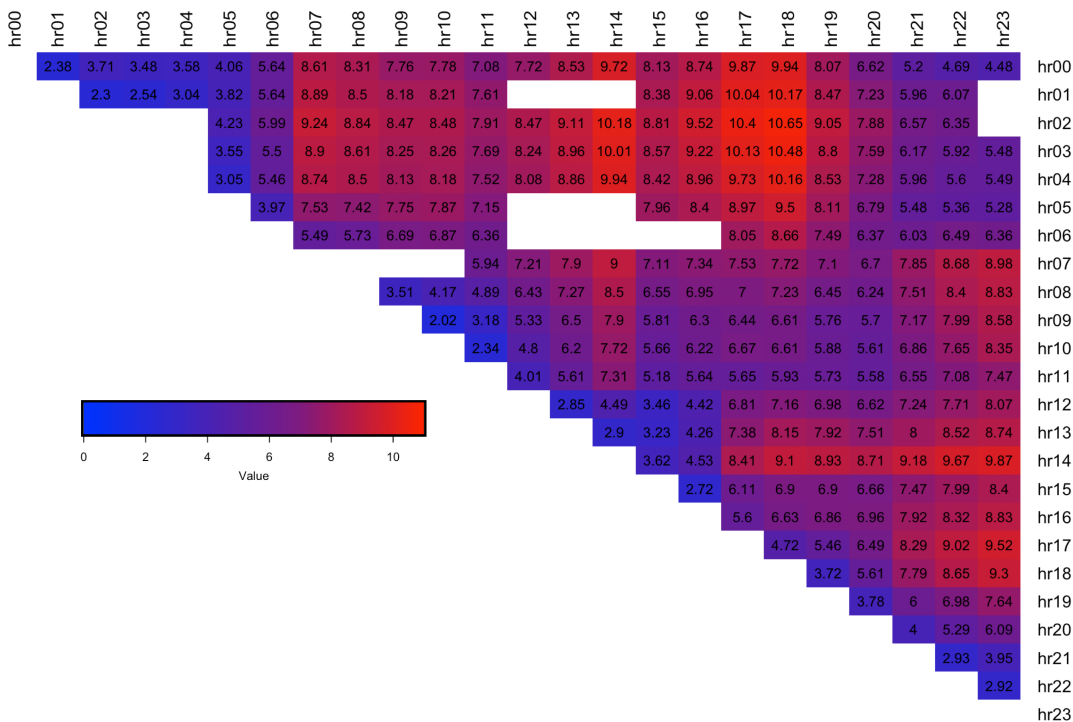

**Figure 23.** RMSE of Normal type distributions for forecasting expected value, $E(Y)$, over rolling window test data.

### 3.2.2. Pinball-Based Evaluation

A more accurate approach would be to use the Pinball Loss function, which was used to compare the accuracy of the full distribution forecast at each time step $t$ of the forecasted time series for each spread. The analysis used Algorithm 2 where instead of six distributions only two distributions are used: $D^{(i)}, i = 1, 2$, where $i = 1$ represented the best-fitting distribution for model $s$, and $i = 2$ is the benchmark Normal distribution. The estimation of models was not always convergent within the GAMLSS estimation algorithm and certain spreads failed to be fitted either by skewed distribution, by Normal distribution or both, which was mostly due to special characteristics in the dataset. For example, 02–03 hour spread is dominated by noise, with $\frac{1}{3}$ of the data containing 0's, which in turn creates a problem during weight update calculations. If convergence issues were experienced, the forecast for that time step was omitted (which was also accounted for during calculation of Pinball Loss averages). If more than 200 out of 383 forecasts were missing for the best skew distribution of a given spread, the result was judged as unreliable and the forecasts for that spread were treated as unavailable. This happened for 4 out of 276 spreads: hours 02–03/12; 03–05; 13–14. The number of times at least 1 forecast was missing for skew type distribution was 46 times (at most 70 times out of 383 time steps). The convergence issue was also experienced while using the Normal distribution (benchmark), however, forecasts with more than 200 time steps were observed more frequently with a total of 13 out of 276 spreads: hours: 01–12/13/14/23; 02–03/23; 05–13; 06–12/13/14/15/16; 07–08. When this happened the skew type distribution was judged to have the better result. The number of times at least 1 forecast was missing for Normal type distribution was 5 times (at most 42 time steps out of 383). The Pinball Loss score $\bar{\mathbf{L}}$ was calculated by averaging the Pinball Loss $\mathbf{L}(q_a, y_t)$ (Equation (11)) across all $t$ rounded to 5 decimal places. If forecast at $t$ was missing due to a failed model, and if total missing values over a forecasted time-series was less than 100 for a given spread, the PL average was adjusted to the relevant number of remaining forecasts.

The results show that for 276 spreads the skewed distributions forecasted the full density more accurately 258 times vs. 18 times for the Normal distribution, when PL score is calculated to 5 decimal places. Reducing decimal places to 3, results in the same number of Normal distributions (18 better than Skew), however, in this case results show that for 5 spreads a Normal distribution produces as good a fit as the skew distribution, and 253 cases of where skew is a better distribution for spreads. The differences between Pinball scores are plotted in Figure 24, where negative values indicate that a skew type distribution has a better forecasting power over Normal. The overwhelming majority preference for skewed distributions is evident through negative values, as expected due to highly skewed and kurtotic nature of spread price data. Specifically, the results show that in 12 cases the skew distribution is strongly better than Normal (dark blue), 90 cases moderately better (navy blue), with the remaining 151 cases where skew is better than Normal.

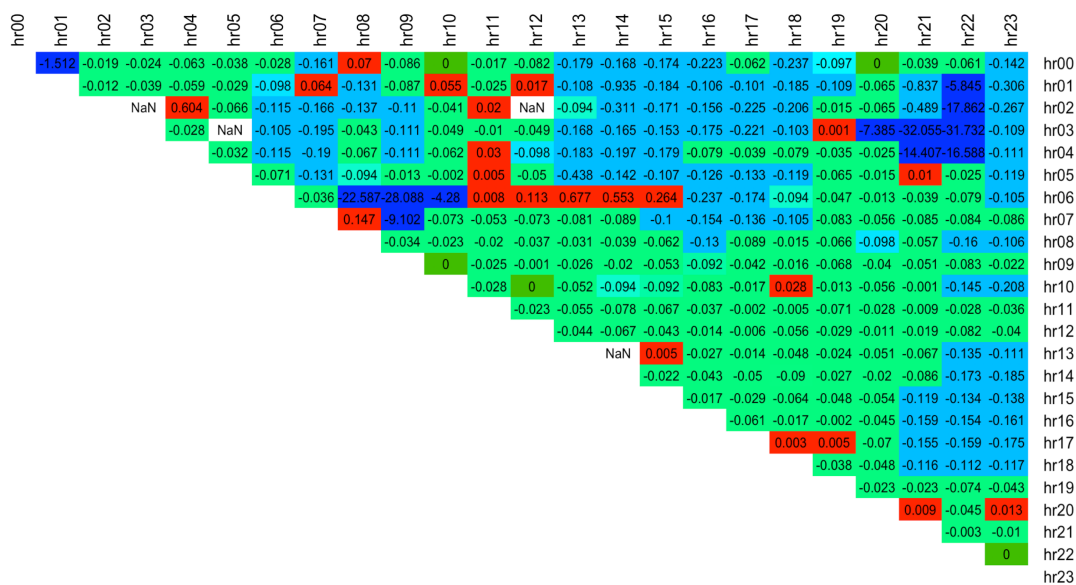

**Figure 24.** Forecasting power of skew vs. Normal distributions using a difference of Pinball Loss scores (negative sign means skew type distribution outperforms Normal).

### 3.2.3. Diebold–Mariano-Based Evaluation

To draw statistically significant conclusions over the outperformance of skew type distribution forecast accuracy over that of the Normal distribution, the Diebold–Mariano [15] test is performed. The test is applicable to forecast errors that do not have zero mean, that are not Gaussian, and that may be serially/contemporaneously correlated, thus Diebold–Mariano accounts for the correlation structure between the errors. The test is valid for cases where a non-quadratic loss function is used, such as the Pinball Loss function used here. We use a variation of the standard Diebold–Mariano test with the implementation proposed in [16]. For each spread number $s$ we have $t = 1, ..., 383$ forecasts produced by two models ($M_1$—estimated with skew type distribution and $M_2$—estimated with Normal distribution), which are tested against each other using a one-sided test at 5% significance level. The null hypothesis is that the two methods have the same forecast accuracy, with a one-sided alternative hypothesis that forecasting power by skew type distribution outperforms that of the Normal, $H_0 : E(\Delta_{M_1,M_2,t,s}) \leq 0$, where the Loss Differential Series $\Delta_{M_1,M_2,t,s}$ is calculated using

$$\Delta_{M_1,M_2,t,s} = |\bar{L}_{M_1,t,s}| - |\bar{L}_{M_2,t,s}| \tag{14}$$

where $s$ is the spread number, $t$ is the forecast time step, $\bar{L}_{t,1}$ average quantile score of skew type distribution model ($M_1$) forecast at step $t$, $\bar{L}_{t,2}$ average quantile score of Normal distribution ($M_2$) forecast at step $t$.

The $p$-values are displayed in Figure 25 and the results show that out of 276 spreads: (a) the skew type distribution is significantly better at forecasting the spreads 161 times at 5% (bright green) and 15 times at 10% (olive green). Note: the distributions which were used to learn these 176 models break into JSU: 9 times, SEP1: 14 times, SEP2: 16 times, ST1: 13 times, ST2: 18 times and ST5: 106 times ($\frac{106}{164} \times 100 = 64.5\%$); (b) the Normal and the skew have the same forecasting power 48 times (i.e., the null hypothesis could not be rejected at 10%). Note: the distributions which were used to learn these 48 models break into JSU: 9 times, SEP1: 4 times, SEP2: 8 times, ST1: 8 times, ST2: 4 times and ST5: 15 times ($\frac{15}{48} \times 100 = 31.25\%$); (c) the results could not be obtained for the skew distribution model 49 times (white spaces of upper triangular) since at least 1 forecast was missing due to Quantile estimate convergence issues. Note: the distributions which were used to learn these 49 models break into JSU: 4 times, SEP1: 8 times, SEP2: 21 times, ST1: 9 times, ST2: 5 times and ST5: 2 times ($\frac{2}{49} \times 100 = 4\%$) (i.e., this typically happened for distributions other than ST5). Note: the Normal distribution approach may have resulted in missing forecasts also, however this was not checked.

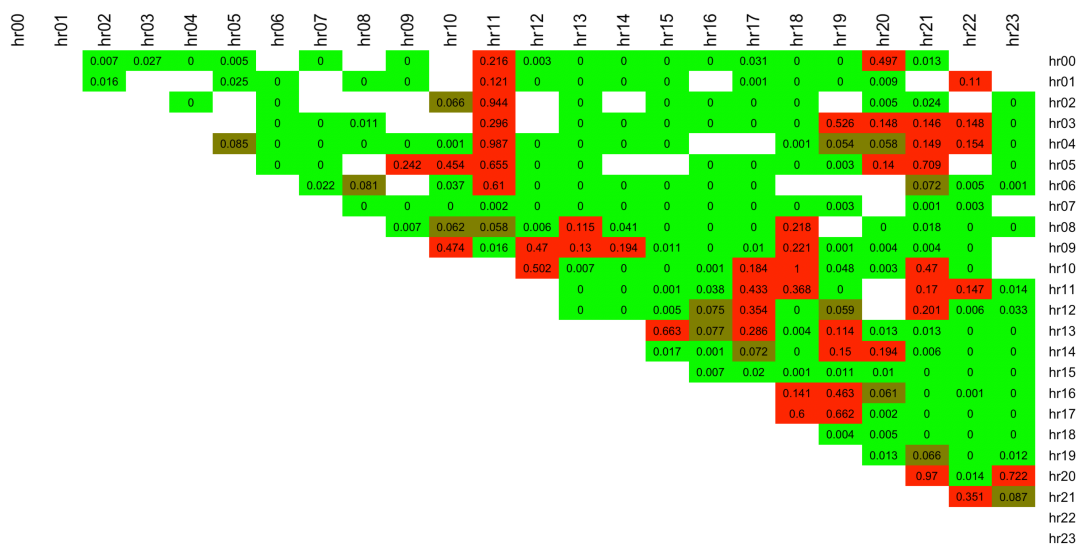

**Figure 25.** Diebold-Mariano test $p$-values (green at 5% significance, olive at 10% significance, red insignificant).

We focused on selecting the skew distributions which fitted best the individual spread hour data. This is the most accurate approach, however, some distributions experienced issues with convergence during quantile estimation. As a result of the above analysis it was found that: (a) the ST5 distribution was most frequently selected as the best distribution by simple distribution fit based on $y \sim 1$ and factor-based distribution fit based on both RMSE and Pinball Loss function; (b) the ST5 distribution was the most reliable for convergence of quantile estimates using GAMLSS function qFUN (e.g., qST5), especially for the extreme quantiles of $q_1, q_2, q_3, q_{97}, q_{98}, q_{99}$ where other distributions such as SEP1 and SEP2 often failed. Note that quantile estimates are important for this research because they are utilised in the statistical testing of performance of two distributions. For example, the Pinball Loss calculations revealed that for 49 spread hours at least one forecast step (out of 383 steps) failed to extract 95 quantiles from the estimated model. Upon further examination, it was revealed that out of the 49 occurrences only 4% had ST5 as the underlying distribution for which the model was estimated. This supports our claim that ST5 is a reliable distribution for the quantile estimates; (c) the best distribution for each spread was chosen based on the Pinball Loss score calculated for the validation data. Each spread had 6 possible distributions from which the best distribution was chosen and the one with the lowest score was taken as the best distribution for that spread. Further analysis reveals that on 1/3 of occasions when other distributions than ST5 were selected as 'best', the ST5 was the second-best distribution, which on average was only worse by 1.44%. However, for the spreads where the ST5 was the best distribution, the pinball loss score was on average better by 4.84%, which points to the possibility that the best distribution did not have a significantly different performance when compared to ST5. Therefore we conclude that if one wished to use a single distribution across all spread hours, the ST5 distribution forms a robust choice. However, our research shows that for a more rigorous analysis the more detailed approach should be used, where individual spreads have distributions of best fit assigned to them.

## 4. Conclusions

The methodology developed in this paper is innovative in focusing directly upon modeling the intraday spreads in day ahead prices, rather than the day ahead price levels themselves. The density function for each spread is modeled directly. As the day ahead hourly prices are non-Normal and not independent, it is not generally possible to derive the spread densities from pairs of price densities. This work is motivated by potential applications to short-term arbitrage trading and battery operations with risk-averse traders, but we have restricted this research to considerations of the forecastability of these spreads. In this context, the value of detailed, computationally intensive modeling of intra-day power price spread densities using a flexible four-parameter distributional form, generally the skew-t. This allows the dynamic conditional parameter estimates to follow stochastic evolutions driven by exogenous factors, most importantly the day ahead demand, wind and solar forecasts. The result is a dynamic specification of a four-parameter density function, adapting its shape to the evolution of demand, wind and solar forecasts. Furthermore, the analytical properties of these four parameter functions allow direct computation of their cumulative distribution functions. This is an attractive alternative to conventional quantile regression methods that require interpolations between a discrete set of quantile estimates to approximate the full distribution function.

With respect to validation and out-of-sample testing, these forecasts performed well in a rolling window backtesting and outperformed baseline comparison to a Normal density model. The significance of outperformance was established, generally at 5%, with Diebold–Mariano tests on the Pinball Loss functions for the densities. Insofar as the Normal baseline was outperformed, this is a validation of the approach against the simpler alternative. However, the model specification and validation process are computationally intensive, and whilst modeling simplifications could be introduced, generally, we think extra accuracy is important. Overall, this formulation and application shows the merits of a computationally intensive approach to accurate specifications and these are likely to be more attractive in practical applications, e.g. to intraday arbitrage trading, than methods based upon analytical simplifications of the stochastic price processes. The value of this methodology should be quite general to most electricity markets where spreads are relevant for short-term, intraday arbitrage trading. The key ingredient is the non-Normality of the prices and the dependence of the price density shapes on potential exogenous variables. With the widespread introduction of wind and solar production facilities, these conditions are likely to be fulfilled in most daily power markets.

**Author Contributions:** Both authors contributed equally in all stages of the research. All authors have read and agreed to the published version of the manuscript.

**Funding:** This research received no external funding.

**Conflicts of Interest:** The authors declare no conflict of interest.

## Appendix A. Data Collection and Pre-Processing

The data collection process for each variable is described below.

### Appendix A.1. German Day-Ahead Electricity Price

Day-ahead prices for each hour of the day ahead was downloaded from http://www.energinet.dk/en/el/engrosmarked/udtraek-af-markedsdata/Sider/default.aspx.

### Appendix A.2. Total Wind Day Ahead Forecast

The total wind day ahead forecasts were obtained as the sum of wind forecasts for the four control areas, each downloaded from https://data.open-power-system-data.org. Missing data for two control areas of the following dates was interpolated: (a) 50 Hertz (MW): 1 January 2017–31 March 2017 hourly wind forecast data replaced with average of four 15 min intervals starting on the hour, e.g. for data for hour 23.00 is the average of 23:00–23:15, 23:15–23:30, 23:30–23:45, 23:45–00:00 obtained from http://www.50hertz.com/en/Grid-Data/Wind-power/Archive-Wind-power. (b) TransnetBW (MW): the data for the following dates was missing (and also missing from https://www.transnetbw.com/en/transparency/market-data/key-figures) 23 March 2012; 12 October 2012; 21 February 2013; 31 March 2013; 30 March 2014; 7 September 2014; 29 March 2017 and thus were replaced with an average of the same hour from the past 7 days. Note the March dates experience a change in the clock, therefore for consistency of data the average (for each hour for end of March missing data) was shifted by one hour: i.e., for hour 06, an average of 7 previous days of hour 05 was taken instead of 06.

### Appendix A.3. Total Solar Day Ahead Forecast

Each control area's forecast is obtained by finding the average of 15 min forecasts starting from the hour (e.g., for 23.00 data point this is the average of 23:00–23:15, 23:15–23:30, 23:30–23:45, 23:45–00:00), with data downloaded from https://data.open-power-system-data.org. The total solar day-ahead forecast is found by summing the average for each of the four control regions. A number of days contained non-zero entries for forecasted sunshine during hours 22.00 and 23.00 in the night (e.g., 30 June 2014 contained entries 224 and 15 respectively). In order aid smooth regression estimation, any non-zero entries for hours 22.00/23.00 have been overwritten with 0 in order to be consistent with the majority of the data for those hours (approx. 10 entries in total). Note: the processed data set therefore did not contain any forecasted sunshine for hours: 22.00, 23.00, 24.00, 00.00, 01.00, 02.00, 03.00. Some data was missing and was therefore replaced for the following control areas: (a) 50 Hertz (MW): 1 January 2017–31 March 2017 hourly data replaced with solar forecast, average of four 15 min intervals starting on the hour: e.g., for 23.00 data is the average of 23:00–23:15, 23:15–23:30, 23:30–23:45, 23:45–00:00 (data downloaded from http://www.50hertz.com/en/Grid-Data/Photovoltaics/Archive-Photovoltaics) 13 May 2014 (original data from the same website suggests a forecast of full sunshine at 00 midnight to 23.00 of 2000+ MW; it was therefore replaced with an average of last 7 days of forecasts for each hour). (b) Amprion (MW): Date 28 April 2014 was missing thus hourly data was replaced with solar forecast, average of four 15 min intervals starting on the hour: e.g., for 23.00 data is the average of 23:00–23:15, 23:15–23:30, 23:30–23:45, 23:45–00:00 (data downloaded from http://www.amprion.net/en/photovoltaic-infeed). (c) TransnetBW (MW): The days 31 March 2013; 30 March 2014; 7 September 2014; 29 March 2015 were missing from https://www.transnetbw.com/en/transparency/market-data/key-figures and thus were replaced with an average of the hour from the past 7 days. Note the March dates experience a change in the clock, therefore for consistency of data, the average (for each hour for end of March missing data) was shifted by one hour: i.e., for hour 06, an average of 7 previous days of hour 05 was taken instead of 06. Incorrect data for the 13 December 2014 data shows sunshine at night and for the 24 hours of that day (200+ MW), hence this was replaced with an average of last 7 days of forecasts for each hour.

*Appendix A.4. Day Ahead Total Load*

Data was downloaded from https://transparency.entsoe.eu/load-domain/r2/totalLoadR2/show containing 2 different total load variables: 'load old' and 'load new' (overlapping from year 2015 only). We used 'load old' data from 1 January 2012 to 31 December 2015 and 'load new' data from 1 January 2016 to 31 March 2017. Each hourly total load data point was created (by the data provider) as the sum of the average of 15min segments for each control area. For example, load of hour 23.00 was calculated as the sum of the average of segments 23:00–23:15, 23:15–23:30, 23:30–23:45, 23:45–00:00 for each control region.

The forecasted load (day ahead total load) was calculated by summing up total load values for hours 00.00–10.00 from same day and total load for hours 11.00–23.00 from the day before. The logic for this accumulation is as follows: in order to submit a bid for tomorrow a trader would need to submit all data by 12.00 noon, therefore they would use the realised total load from today's hours starting from 10.00 am and work backwards to yesterday 11.00 am. Since 11 am is calculated as the average load for $4 \times 15$ min segments starting from 11.00 am and finishing at 12.00, it is most likely that 12.00 would not be met, thus an 11.00 cut off is suggested.

*Appendix A.5. Steam Coal ARA 1 Month Forward*

Benchmark index for steam coal delivered to the Amsterdam Rotterdam Antwerp region of the Northwest Europe (ARA), quoted as a single price per day. Original Source: Bloomberg ticker API21MON.

*Appendix A.6. Gas Day-Ahead Forward*

Germany Gaspool (GPL) natural gas day-ahead forward (delivered next working day), quoted as a single price per day. Original Source: Bloomberg ticker BHDAHD.

*Appendix A.7. Dummy*

The weekly seasonality and holidays are included into a single dummy variable, which takes on value 1 for Saturday/Sunday and the following German state holidays: New Year's Day, Good Friday, Easter Monday, Labour Day, Ascension Day, Whit Monday, German unit day, Christmas Day, Boxing day and New Years Eve. The following holidays have been omitted because of the focus on German TSO market: Austrian public holidays/regional holidays, such as Pentecost, Corpus Christi Assumption of Mary, Reformation day, All Saints day and Repentance Day.

## Appendix B. Distributions

### *Appendix B.1. SEP1 Distribution*

The probability density function of skew exponential power type 1 distribution [17] is given by

$$f_Y(y_t|\mu_t, \sigma_t, \nu_t, \tau_t) = \frac{2}{\sigma_t} f_{Z_1}(z_t) F_{Z_1}(\nu_t z_t)$$

where $Z_1 \sim PE2(0, \tau_t^{1/\tau_t}, \tau_t)$ and $PE2$ is the exponential type 2 distribution.

The expected value of the random variable $Y_t$ is given by

$$E(Z_t) = \text{sign}(\nu_t)\tau_t^{1/\tau_t} \frac{\Gamma(\frac{2}{\tau_t})}{\Gamma(\frac{1}{\tau_t})} pBEo\left(\frac{\nu_t^{\tau_t}}{1+\nu_t^{\tau_t}}, \frac{1}{\tau_t}, \frac{2}{\tau_t}\right) \quad (A1)$$

where $\Gamma(\cdot)$ is the Gamma function, $pBEo(q, a, b)$ is the cumulative distribution function of the beta distribution $BEo(a, b)$, evaluated at $q$, with shape parameters $a$ and $b$. Note: certain values of $\nu_t$ and $\tau_t$, resulted in a numerical error when calculating the exponent, $\nu_t^{\tau_t}$, and whenever this happened the value of this quantity was set to 1.

### *Appendix B.2. SEP2 Distribution*

The probability density function of skew exponential power type 2 distribution [17,18] is given by

$$f_Y(y_t|\mu_t, \sigma_t, \nu_t, \tau_t) = \frac{2}{\sigma} f_{Z_1}(z_t) \Phi(\omega_t)$$

where $\omega_t = \text{sign}(z_t)|z_t|^{\tau_t/2}\nu_t\sqrt{\frac{2}{\tau_t}}$, $Z_1 \sim PE2(0, \tau_t^{1/\tau_t}, \tau_t)$, $\Phi(\omega_t)$ is the cdf of the standard normal random variable evaluated at $\omega_t$.

The expected value of the random variable $Y_t$ is given by

$$E(Z_t) = \frac{2\tau_t^{1/\tau_t}\nu_t}{\sqrt{\pi}\tau_t\Gamma(\frac{1}{\tau_t})(1+\nu_t^2)^{2/\tau_t+0.5}} \sum_{n=0}^{\infty} \frac{\Gamma(\frac{2}{\tau_t}+n+0.5)}{(2n+1)!!}\left(\frac{2\nu_t^2}{1+\nu_t^2}\right)^n \quad (A2)$$

where $(2n+1)!! = 1.3.5...(2n-1)$. Note, when calculating the summation term, we set a cap of $n = 10$ iterations which was found to provide sufficient convergence.

### *Appendix B.3. ST2 Distribution*

The probability density function of the skew t type 2 distribution [19] is given by

$$f_Y(y_t|\mu_t, \sigma_t, \nu_t, \tau_t) = \frac{2}{\sigma} f_{Z_1}(z_t) F_{Z_2}(w_t)$$

where $w_t = \nu_t \lambda_t^{0.5} z_t$, $\lambda_t = \frac{\tau_t+1}{\tau_t+z_t^2}$, $Z_1 \sim TF(0, 1, \tau_t)$, $Z_2 \sim TF(0, 1, \tau_t+1)$, $TF(\cdot)$ is a t distribution with $\tau_t > 0$ degrees of freedom.

The expected value of the random variable $Y_t$ is given by

$$E(Z_t) = \frac{\nu_t \tau_t^{0.5}\Gamma(\frac{\tau_t-1}{2})}{\pi^{0.5}(1+\nu_t^2)^{0.5}\Gamma(\frac{\tau_t}{2})} \quad (A3)$$

where $\tau_t > 1$. Note, the expected value $E(Z_t)$ was set to 0 any time that $(\tau_t - 1) < \text{tol}$, where tol was set to 0.05, in order to limit large outputs from $\Gamma(n)$ when $n \to 0$.

*Appendix B.4. ST5 Distribution*

The probability density function of the skew t type 5 distribution [20] is given by

$$f_Y(y_t|\mu_t, \sigma_t, \nu_t, \tau_t) = \frac{c_t}{\sigma_t} \left(1 + \frac{z_t}{(a_t + b_t + z_t^2)^{0.5}}\right)^{a_t + 0.5} \left(1 - \frac{z_t}{(a_t + b_t + z_t^2)^{0.5}}\right)^{b_t + 0.5}$$

where $c = \left(2^{a_t + b_t - 1}(a_t + b_t)^{0.5} B(a_t, b_t)\right)^{-1}$, $\nu_t = \frac{(a_t - b_t)}{(a_t b_t (a_t + b)_t)^{0.5}}$, $\tau_t = \frac{2}{(a_t + b_t)}$, $B(a_t, b_t)$.

The expected value of the random variable $Y_t$ is given by

$$E(Z_t) = \frac{(a_t - b_t)(a_t + b_t)^{0.5} \Gamma(a_t - 0.5)\Gamma(b_t - 0.5)}{2\Gamma(a_t)\Gamma(b_t)} \tag{A4}$$

Note: this corrects a typo in the specification of $E(Z_t)$ of the Manual [12,21], which specifies $\Gamma(a_t - 0.5)\Gamma(a_t - 0.5)$.

Once the parameters $\nu_t$ and $\tau_t$ are estimated, re-arranging the system of two equations (see Appendix B.5 for derivation) results in the following expressions for $a_t$ and $b_t$

$$a_t = \frac{2}{\tau_t} - b_t \tag{A5}$$

$$\nu_t = \frac{2(1 - b_t \tau_t)}{\left(2b_t(2 - b_t \tau_t)\right)^{0.5}} \tag{A6}$$

where $b_t$ is found first using Newton-Raphson root searching algorithm with the R function `uniroot` (for certain values of $\nu_t, \tau_t$ the root of this function does not exist, in this case we set $b_t = 0$). Note: In order to limit large outputs from $\Gamma(n)$ when $n \to 0$, a restriction was imposed on values of $a_t, b_t$ to be positive and for $(a_t - 0.5) <$ tol, $(b_t - 0.5) <$ tol, where tol $= 0.05$, and if this condition was not met we set $E(Z_t) = 0$.

*Appendix B.5. ST5 System of Equations*

The parameters $a$ and $b$ are derived by rearranging the system of two equations, where $\nu$ and $\tau$ are known.

$$\nu = \frac{(a - b)}{\left(ab(a + b)\right)^{0.5}} \tag{A7}$$

$$\tau = \frac{2}{(a + b)} \tag{A8}$$

Hence, by re-arranging Equation (A8), $a$ equals:

$$a = \frac{2}{\tau} - b \tag{A9}$$

Substituting $a$ into Equation (A7) we obtain

$$\nu = \frac{\frac{2}{\tau} - b - b}{\left[ b \left( \frac{2}{\tau} - b \right) \left( \frac{2}{\tau} - b + b \right) \right]^{0.5}}$$

$$\nu = \frac{\frac{2}{\tau} - 2b}{\left[ b \left( \frac{2}{\tau} - b \right) \left( \frac{2}{\tau} \right) \right]^{0.5}}$$

$$\nu = \frac{\frac{2(1 - b\tau)}{\tau}}{\left[ b \left( \frac{2 - b\tau}{\tau} \right) \frac{2}{\tau} \right]^{0.5}}$$

$$\nu = \frac{\frac{2(1 - b\tau)}{\tau}}{\left[ 2b \left( \frac{2 - b\tau}{\tau} \right) \frac{1}{\tau} \right]^{0.5}}$$

$$\nu = \frac{2(1 - b\tau)}{\left[ 2b(2 - b\tau) \right]^{0.5}} \tag{A10}$$

Equation (A10) is re-arranged to have the form $f(x) = 0$ (see Equation (A11)) and solved for $b$ using Newton-Raphson root searching algorithm.

$$0 = \frac{2(1 - b\tau)}{\left[ 2b(2 - b\tau) \right]^{0.5}} - \nu \tag{A11}$$

## Appendix C. Training Stage

### Appendix C.1. Spreads—Best Distribution Fit (Factor-Based)

The following spread hours were not learnt for distributions:
**JSU** 21–22. **ST2** 02–03. **ST5** 01–02; 02–03.
**SEP1** 00–02; 00–20; 01–02; 01–09; 01–17; 01–21; 01–22; 02–03; 02–04; 02–05; 02–09; 02–10; 02–17; 02–20; 02–21; 02–23; 02–07; 03–08; 03–21; 04–07; 04–10; 04–21; 05–07; 05–09; 05–21; 05–22; 06–07; 06–08; 06–09; 06–10; 06–17; 06–20; 06–22; 07–08; 07–09; 07–10; 12–14; 13–14; 13–21; 21–23.
**SEP2** 01–02; 01–07; 01–21; 02–03; 02–04; 02–05; 02–18; 02–21; 02–22; 03–04; 03–07; 03–21; 04–07; 04–21; 05–07; 05–09; 05–21; 06–09; 06–10; 06–22; 07–08; 07–10; 07–21; 07–22; 12–13; 13–21; 21–22.
**ST1** 01–02; 02–03; 02–06; 05–21; 06–20; 13–21.

### Appendix C.2. Quantile Estimation

The number of times quantiles were not obtained for distribution $D^i, i = 1, .., 6$ during the validation stage are given in Table A1. The table shows that ST1 and ST5 distributions struggled to extract quantiles the most.

**Table A1.** Number of times for which quantile estimates at $t$ were not obtained for each distribution.

| JSU | SEP1 | SEP2 | ST1 | ST2 | ST5 |
|-----|------|------|-----|-----|-----|
| 0 | 19 | 4 | 176 | 147 | 0 |

## References

1. Weron, R. Electricity price forecasting: A review of the state-of-the-art with a look into the future. *Int. J. Forecast.* **2014**, *30*, 1030–1081. [CrossRef]
2. Nowotarski, J.; Weron, R. Recent advances in electricity price forecasting: A review of probabilistic forecasting. *Renew. Sustain. Energy Rev.* **2017**, *81*, 1548–1568. [CrossRef]
3. Nowotarski, J.; Weron, R. Computing electricity spot price prediction intervals using quantile regression and forecast averaging. *Comput. Stat.* **2015**, *30*, 791–803. [CrossRef]

4.  Garcia-Martos, C.; Rodrıguez, J.; Sanchez, M. Forecasting electricity prices by extracting dynamic common factors: Application to the Iberian market. *IET Gener. Transm. Distrib.* **2012**, *6*, 11–20. [CrossRef]

5.  Karakatsani, N.V.; Bunn, D.W. Fundamental and behavioural drivers of electricity price volatility. *Stud. Nonlinear Dyn. Econom.* **2010**, *14*, 4. [CrossRef]

6.  Wang, L.; Zhang, Z.; Chen, J. Short-term electricity price forecasting with stacked denoising autoencoders. *IEEE Trans. Power Syst.* **2016**, *32*, 2673–2681. [CrossRef]

7.  Jónsson, T.; Pinson, P.; Madsen, H.; Nielsen, H. Predictive densities for day-ahead electricity prices using time-adaptive quantile regression. *Energies* **2014**, *7*, 5523–5547. [CrossRef]

8.  Denholm, P.; O'Connell, M.; Brinkman, G.; Jorgenson, J. *Overgeneration from Solar Energy in California. A Field Guide to the Duck Chart*; National Renewable Energy Laboratory (NREL): Golden, CO, USA, 2015.

9.  Gianfreda, A.; Bunn, D.W. A stochastic latent moment model for electricity price formation. *Oper. Res.* **2018**, *66*, 1189–1203. [CrossRef]

10. Stasinopoulos, D.M.; Rigby, R.A. Generalized additive models for location scale and shape (GAMLSS) in R. *J. Stat. Softw.* **2007**, *23*, 1–46. [CrossRef]

11. Rigby, R.A.; Stasinopoulos, M.D. Mean and dispersion additive models. In *Statistical Theory and Computational Aspects of Smoothing*; Springer: Berlin, Germany, 1996; pp. 215–230.

12. Stasinopoulos, M.; Rigby, R.; Akantziliotou, C. Instructions on How to Use the GAMLSS Package in R. 2014. Available online: http://www.gamlss.org/ (accessed on 23 January 2020).

13. Bishop, C.M. *Pattern Recognition and Machine Learning*; Springer: Berlin, Germany, 2006.

14. Maciejowska, K.; Nowotarski, J. A hybrid model for GEFCom2014 probabilistic electricity price forecasting. *Int. J. Forecast.* **2016**, *32*, 1051–1056. [CrossRef]

15. Diebold, F.X.; Mariano, R.S. Comparing predictive accuracy. *J. Bus. Econ. Stat.* **2002**, *20*, 134–144. [CrossRef]

16. Harvey, D.; Leybourne, S.; Newbold, P. Testing the equality of prediction mean squared errors. *Int. J. Forecast.* **1997**, *13*, 281–291. [CrossRef]

17. Azzalini, A. A class of distributions which includes the normal ones. *Scand. J. Stat.* **1985**, *12*, 171–178.

18. DiCiccio, T.J.; Monti, A.C. Inferential aspects of the skew exponential power distribution. *J. Am. Stat. Assoc.* **2004**, *99*, 439–450. [CrossRef]

19. Azzalini, A.; Capitanio, A. Distributions generated by perturbation of symmetry with emphasis on a multivariate skew t-distribution. *J. R. Stat. Soc. Ser.* **2003**, *65*, 367–389. [CrossRef]

20. Jones, M.; Faddy, M. A skew extension of the t-distribution, with applications. *J. R. Stat. Soc. Ser.* **2003**, *65*, 159–174. [CrossRef]

21. Rigby, R.; Stasinopoulos, D. *A Flexible Regression Approach Using GAMLSS in R*; London Metropolitan University: London, UK, 2009.